\documentclass[review]{elsarticle}

\usepackage{multirow}
\usepackage[colorlinks]{hyperref}
\usepackage{lscape}
\usepackage{longtable}
\usepackage{color}
\usepackage{graphicx}
\usepackage[labelformat=simple]{subcaption}

\DeclareCaptionLabelFormat{subcaptionlabel}{\normalfont(\textbf{#2}\normalfont)}
\usepackage[textwidth=340pt,textheight=550pt]{geometry}

\journal{Energy Conversion and Management}

\begin{document}
	
	\begin{frontmatter}
		
		\title{A graphical approach to carbon-efficient spot market scheduling for Power-to-X applications}
		
		
		\author[mymainaddress]{Neeraj Bokde}
		
		\author[ento]{Bo Tranberg}

		\author[mymainaddress]{Gorm Bruun Andresen\corref{mycorrespondingauthor}}
		\cortext[mycorrespondingauthor]{Corresponding author}
		\ead{gba@eng.au.dk}
		
		\address[mymainaddress]{Department of Engineering - Renewable Energy and Thermodynamics, Aarhus University, 8000, Denmark}
		\address[ento]{Ento Labs ApS, Inge Lehmanns Gade 10, 6, 8000, Aarhus C, Denmark}
		
		\begin{abstract}
In the Paris agreement of 2015, it was decided to reduce the CO$_2$ emissions of the energy sector to zero by 2050 and to restrict the global mean temperature increase to $1.5^\circ$C above the pre-industrial level. Such commitments are possible only with practically CO$_2$-free power generation based on variable renewable technologies. Historically, the main point of criticism regarding renewable power is the variability driven by weather dependence. Power-to-X systems, which convert excess power to other stores of energy for later use, can play an important role in offsetting the variability of renewable power production. In order to do so, however, these systems have to be scheduled properly to ensure they are being powered by low-carbon technologies.
In this paper, we introduce a graphical approach for scheduling power-to-X plants in the day-ahead market by minimizing carbon emissions and electricity costs. This graphical approach is simple to implement and intuitively explain to stakeholders.
In a simulation study using historical prices and CO$_2$ intensity for four different countries, we find that the price and CO$_2$ intensity tends to decrease with increasing scheduling horizon. The effect diminishes when requiring an increasing amount of full load hours per year. Additionally, investigating the trade-off between optimizing for price or CO$_2$ intensity shows that it is indeed a trade-off: it is not possible to obtain the lowest price and CO$_2$ intensity at the same time.					
		\end{abstract}
		
		\begin{keyword}
			CO$_2$ emission \sep Power-to-X \sep demand flexibility
		\end{keyword}
		
	\end{frontmatter}
	
	\section{Introduction}
	
	The world agreed to reduce the net CO$_2$ emissions of the energy sector to zero by 2050 and to restrict the global mean temperature rise under $1.5^\circ$C above the pre-industrial level, as committed in the UN Climate Change Conference, Paris in 2015 \cite{a1}. These kinds of commitments are possible only with practically CO$_2$-free power generation and the energy systems based on solar, wind, and other renewable sources of power \cite{a2, a3}. To achieve such targets, new policies are necessary to handle climate change by reducing the renewable energy cost and the decarbonization of the energy systems \cite{a4, a5}.
	
	In recent decades, large quantities of variable renewable electricity (VRE) based on wind and solar energy has been introduced in the electricity system. Here, a continuous balance between the electricity generated and consumed needs to be maintained for the system to be stable. High shares of VRE in the electricity system can pose a challenge to this balance \cite{a6}. Traditionally, electricity network operators has been relying on backup power from non-renewable and fossil fuels \cite{a7, a8}. The concept of converting VRE to hydrogen gained attention in recent years in order to mitigate the challenges from the intermittent nature of VREs. Power-to-X systems such as Power-to-Gas, Power-to-Chemical, etc. are emerging as an alternative solution for storing renewable power and to produce CO$_2$ neutral alternatives to fossil fuels for aviation, heavy transport and other sectors that are inherently difficult to electrify \cite{a9, a10, a11}.
	
	The predominantly used form of Power-to-X systems is the Power-to-Gas. The Power-to-Gas system is the conversion of the excess of electricity into storable hydrogen through a water electrolysis process. This stored hydrogen can be further converted into methane, electricity, liquid fuels, etc or directly consumed as energy in several applications \cite{a12, a13}. Power-to-X is the utilization of stored hydrogen or methane into several usable energy forms to store electricity forms, e.g., VRE sources. The main applications of Power-to-X systems include converting VREs into easily storable chemical energy carriers, producing fuels for transportation, industry, and households, as well as the production of chemicals for the industry \cite{a14, a15, a16}. There are numerous applications of Power-to-X systems to achieve a carbon-neutral climate, which includes the transport sector by a mix of electric and hydrogen fuel vehicles or the generation of renewable biofuels with Power-to-Liquid systems. Further, Power-to-X systems have been used in biomass-based plants. Such Power-to-X plants could be used to produce chemicals and renewable decarbonized fuels when biomass cannot be utilized or there is a lack of chemicals like hydrogen in bio-process. 
	
	The power-to-X systems can be beneficial for the electricity day-ahead market (spot market). The negative prices of electricity are one of the major challenges in the VRE spot markets \cite{a17} and such cases can be avoided with the integration of the spot market with the Power-to-X systems \cite{a18}. Looking at the benefits and potential to decarbonize the energy system, in recent years, the Power-to-X systems are making headway and practically exercised at industrial scale \cite{a19}. Some of these successful models are discussed in the studies \cite{a20, a21, a22}.
	
	Though Power-to-X systems have great potential, more success can be achieved with proper planning. Recently, analytical and data-driven studies have been proposed to make the Power-to-X systems more successful \cite{a23, a25, a26}. These studies claimed that the Power-to-X plant operations and efficiency can be improved notably with day or week ahead optimized planning \cite{a24}. The planned activities can further improve the economic and commercial viability of Power-to-X plants. The European spot markets are dynamic in nature and hourly electricity prices vary significantly for each season. This is one of the most challenging factors for the commercial viability of the Power-to-X plants. A study based on spot market bidding in Sweden \cite{a18} proposed a bidding strategy based on day-ahead electricity price forecasting and compared it with conventional (without forecasting) bidding approach. The study helped reduce the chances of the purchase of the high-cost electricity and its integration with Power-to-X systems led to control the carbon intensity of the hydrogen.
	
	Whereas existing studies focus on the economic viability of Power-to-X by considering only the market prices, this study considers both market prices and CO$_2$ intensity. This is important to ensure that the power used by the Power-to-X plant is supplied by low-carbon VREs. This guarantees that the flexibility added to the system supports the integration of larger shares of VREs.

	This study proposes a data-driven model for optimal day-ahead spot market scheduling for Power-to-X storage applications. In addition, a graphical model is proposed to achieve inexpensive Power-to-X operation with minimum carbon emissions. The model is proposed, examined, and validated on a real-world time-series, and the effect of operating expenses (OPEX) and capital expenditures (CAPEX) is ignored in it. Consequently, this paper contributes to the literature as follows:
	
	\begin{itemize}
		\item The models to forecast the spot market electricity price and CO$_2$ intensity time-series for 36-48 hours ahead values, which are necessary for electricity market bidding.
		\item Discussing the trade-off between optimizing for spot market electricity price or CO$_2$ intensity.
		\item Optimally scheduling the Power-to-X storage operations by minimizing carbon emissions and electricity costs.
	\end{itemize}

	\section{Methodology}
\begin{figure}[h]
		\centering{\includegraphics[width=\textwidth]{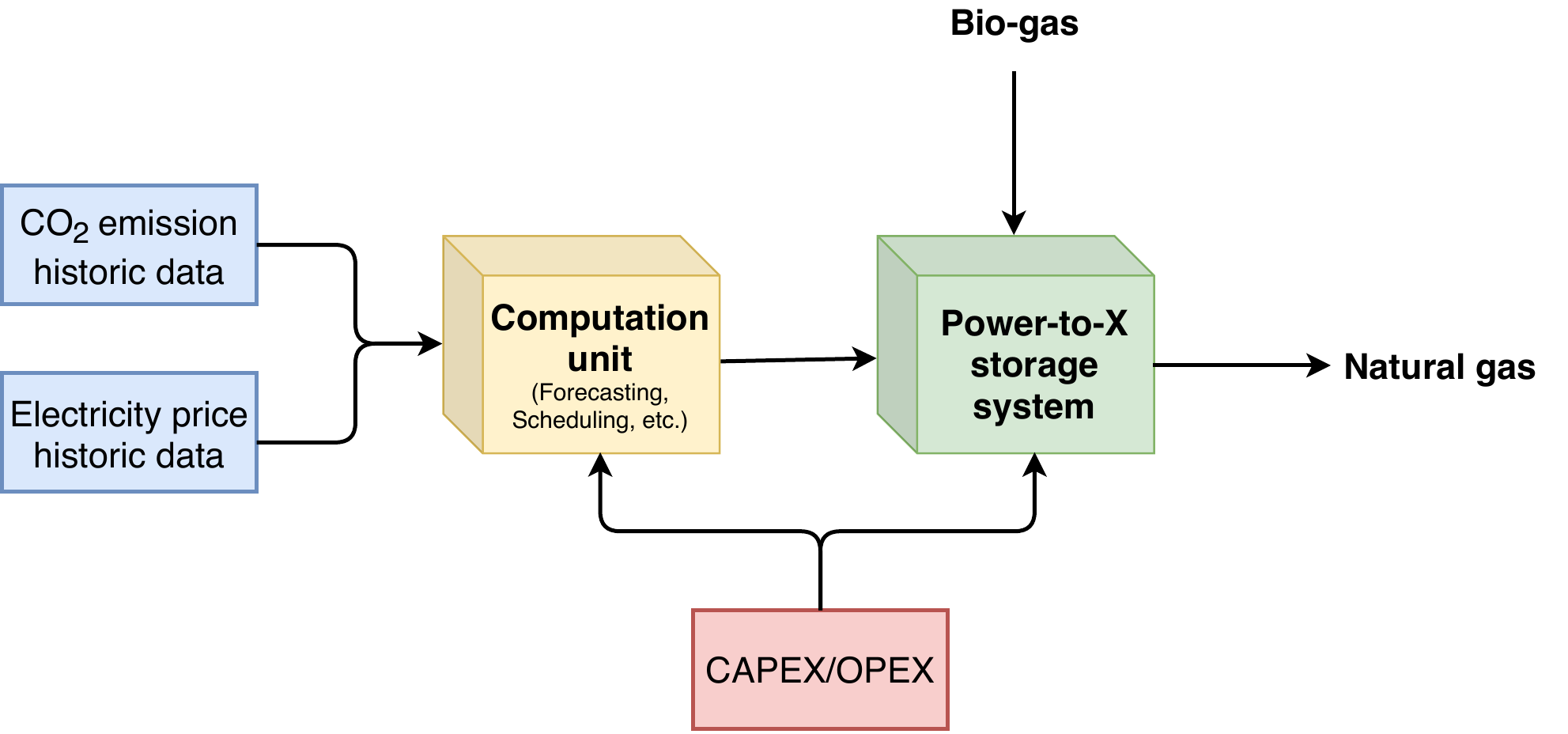}}
		\caption{The framework of the proposed methodology.}
		\label{Fig1}
	\end{figure}

	In the literature, there are several models proposed for spot market scheduling based on the spot market electricity prices, which are directly or indirectly related to Power-to-X systems \cite{b1, b2, b3}. In addition, due to the serious concern towards carbon emission, some recent studies were focused on minimum carbon emission based spot market scheduling \cite{b4}. For the first time, the proposed study is focused on spot market scheduling with the selection of hours having minimum carbon emissions and the least electricity prices for the Power-to-X applications. The framework of the proposed methodology is shown in Figure \ref{Fig1}. The presented model is focused on the optimum day-ahead spot market scheduling for Power-to-X storage systems based on the trade-off of the day-ahead forecasted CO$_2$ emissions and electricity prices. As shown in Figure \ref{Fig1}, the Power-to-X storage system is connected with a computation unit that takes historic electricity price and CO$_2$ intensity datasets for a specific price zone (of a spot market) as an input. The computation unit forecasts both time-series datasets for 36-48 hours-ahead values with suitable forecasting methods. Then optimal scheduling of hours for the storage system is obtained with consideration of the trade-off between the forecasted electricity prices and CO$_2$ intensity values. While performing scheduling, it is assumed that the Power-to-X storage system is owned by a third party (private) operator who wishes to store electricity generated predominantly by VREs with the least carbon emissions and lowest prices and converting it into the natural gases or other forms of fuels and chemicals. The electricity prices and CO$_2$ intensity are the independent parameters and usually, there is no strong correlation between these parameters, hence it is an important step to estimate the trade-off between these parameters and schedule the spot market hours with optimum (least) CO$_2$ emissions and electricity prices. The optimal operation of the Power-to-X storage would therefore responsible for inexpensive and decarbonized sources of energy.

	\subsection{Spot market forecasting}
	Usually, the bidding in the spot market is improving more with proper forecasts, scheduling, and planning, as discussed in recent studies \cite{rad2020optimal, zhao2019multi}. Such studies have mainly focused on the economical aspect of the spot market and electricity usage that eventually used electricity prices as a targeted dataset. However, in the present study, the spot market scheduling is minimizing both CO$_2$ emissions and costs, therefore, both electricity prices and CO$_2$ intensity time-series are used as the input dataset.
	
	The data set used in this study is obtained from the ENSTO-E Transparency platform \cite{transparency}. Individual transmission system operators report data for production mix, power consumption, import and export flows and spot prices to this platform with hourly resolution. In present study, the spot prices are directly used in our simulation and forecasting studies.
	
	The CO$_2$ intensity is calculated using the method of \cite{tranberg2019real}, which introduced a real-time carbon accounting method for the European electricity markets. This method builds on the concept of flow tracing, which maps power flows between importing and exporting countries in an interconnected network \cite{Tranberg2015}. For each area this method considers local production mix, local power consumption as well as imports and exports between neighboring areas. Since imports from a neighboring area depend both on the mix within that area as well as the neighbors from which this area imports, the hourly CO$_2$ intensity is found simultaneously for all areas by solving a set of linear equations as described in \cite{tranberg2019real}. The average CO$_2$ intensity per area per hour, which is used in the following simulation and forecasting studies, is based on a specific CO$_2$ intensity per generation technology per area based on the ecoinvent 3.4 power plant database \cite{ecoinvent}.
	
\begin{figure}[h]
		\centering{\includegraphics[width=\textwidth]{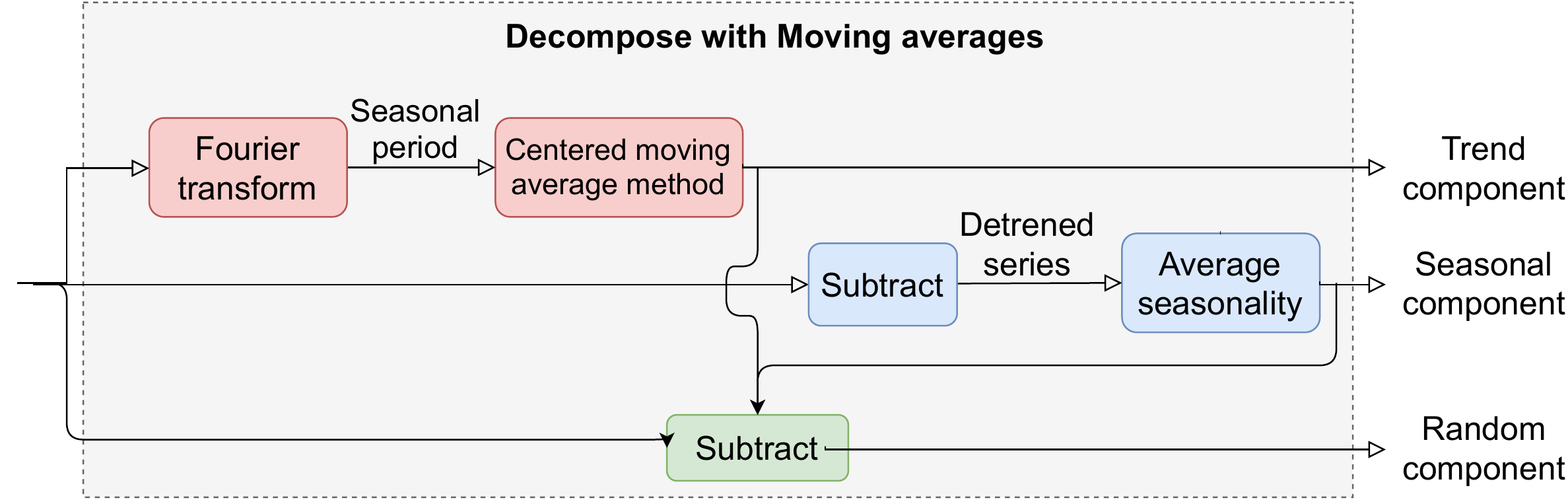}}
		\caption{Schematic of the moving averages based decomposition approaches. (Source: \cite{b4})}
		\label{M1}
	\end{figure}

	
	Spot market scheduling based on forecasted electricity prices is one of the popular research objectives and it can be categorized into short-term forecasting. In the present study, both electricity prices and CO$_2$ intensity are forecasted with the `Method 1' proposed in \cite{b4}. This method is a short-term forecasting model based on the decomposition of the historic time series with moving average method and generation of three sub-series components named: trend, seasonal, and random ones. Figure \ref{M1} shows the block diagram of the moving average based decomposition used in the `Method 1'. The first step of this decomposition in the detection of the trend in the time series using the centered moving average method. Then the time series is detrended by eliminating the trend series from the original time series. The detrended series represents a seasonal component along with some noise values.
	
	The seasonal component is calculated by adding seasonality of detrended series together and dividing it by seasonality period. Finally, with the removal of seasonal and trend components from the original time series, the random noise component is obtained.
	
\begin{figure}[h]
		\centering{\includegraphics[width=3in]{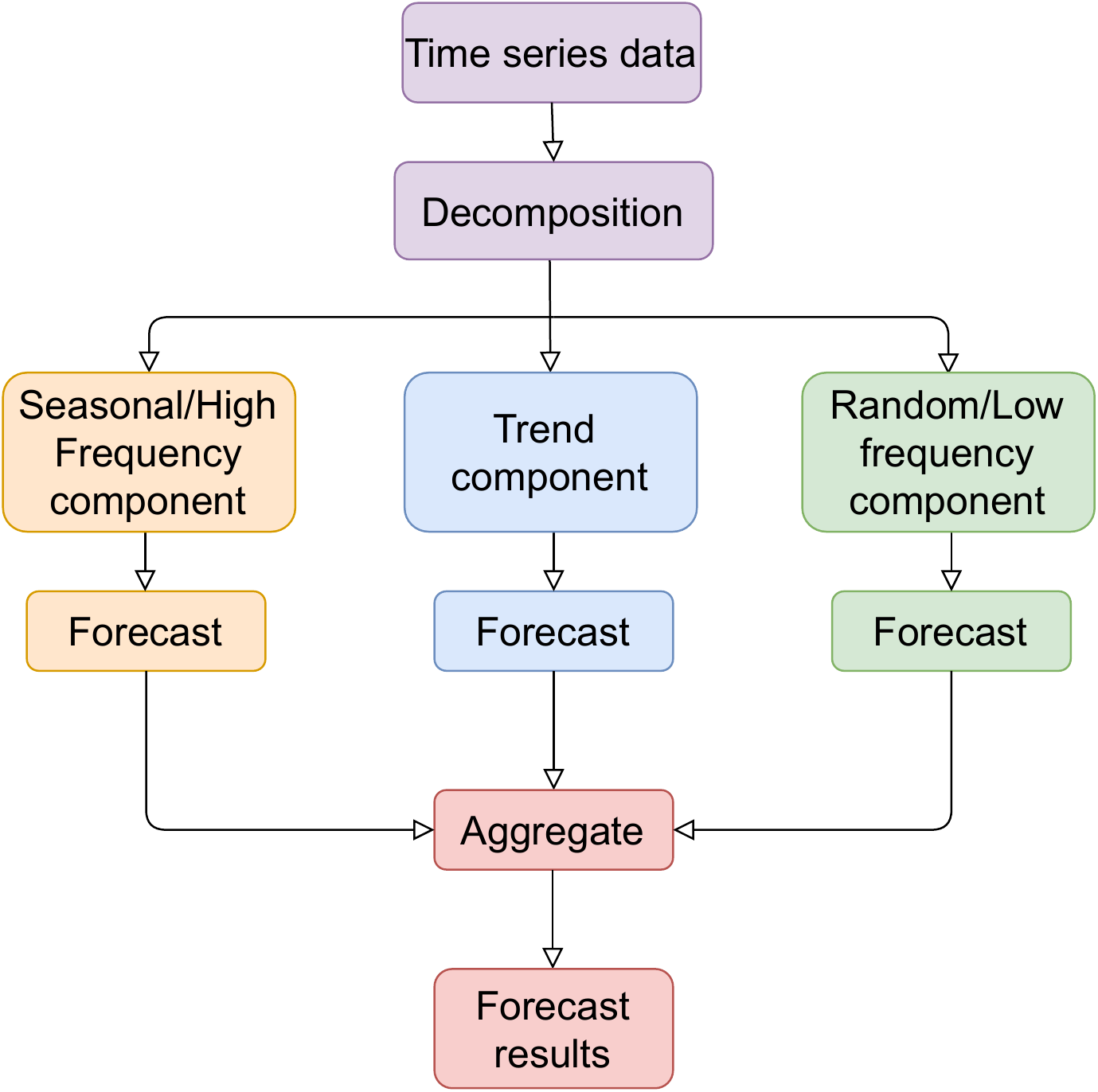}}
		\caption{Block diagram of forecasting method (`Method 1') \cite{b4}.}
		\label{M2}
	\end{figure}

	In `Method 1', all three components are forecasted individually with three different models. The seasonal component is forecasted with Feed-forwards neural network (FFNN) and two different ARIMA models are employed for the trend and random components. The aggregation of the forecasted values of these three components in the final forecasting results as shown in Figure \ref{M2}.
	
	\clearpage
	
	The study proposed in \cite{b4} confirms the higher accuracy of `Method 1' in the short-term CO$_2$ intensity forecasting for several European countries. In the proposed study, `Method 1’ is used to forecast both CO$_2$ intensity and electricity price, and corresponding results are discussed in brief in subsequent section. Though these forecasting results are accurate, it is worth noting that further improvements in forecasting accuracy can lead to more accurate and reliable spot market scheduling.

	\begin{figure}[h]
		\centering{\includegraphics[width=\textwidth]{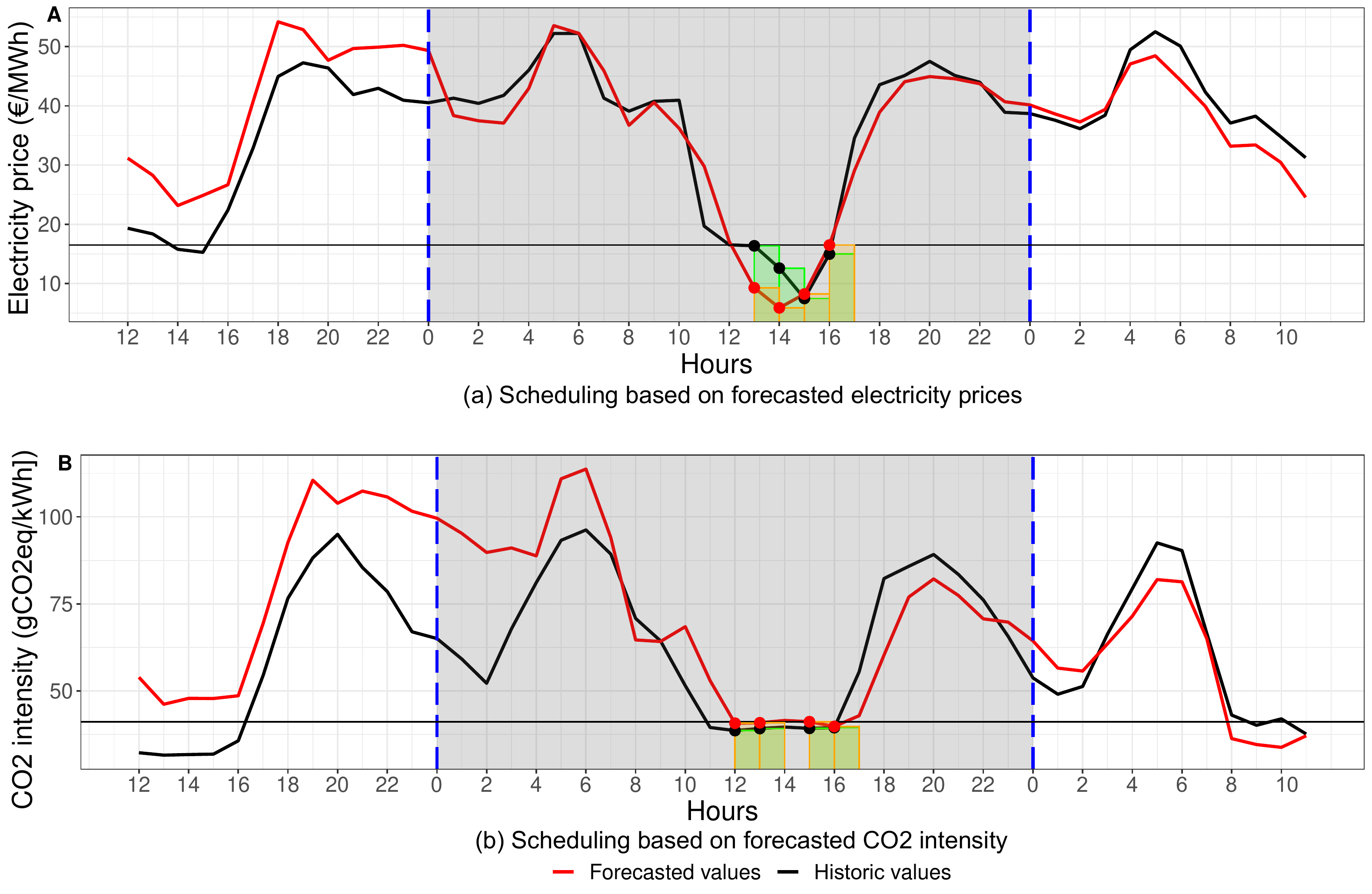}}
		\caption{Example of scheduling 4 hours of flexible electricity consumption one day in
			advance based on (a) electricity prices and (b) CO$_2$ intensity in France. The gray area shows the day-ahead interval. The red and black lines show
			forecasted and realized values of the CO$_2$ intensity, respectively. The four hours minimum
			of the forecast is market with red dots. The corresponding realized values are marked with
			black dots.}
		\label{Fig2}
	\end{figure}
	
	Figures \ref{Fig2} (a) and (b) show the forecasted electricity prices and CO$_2$ intensity time-series with respective forecasting models as discussed in the earlier section. In both sub-figures, the red and black lines represent the forecasted and historic values, respectively. The total forecast horizon shown here is 48-hours and the hours of interest (24 hours) for the bidding purposes are shown in a dark gray background with blue dashed borders. In the European spot markets, bidders need to participate in auctions before noon the day before consumption takes place. Usually, all spot markets open at 10.00 am and auctions end before noon and the bidding can be done for slots (in hours) within 24 hours of the following day, starting from midnight 00.00 am. In both sub-figures, four hours with minimum amount (of CO$_2$ intensity and electricity prices, respectively) are scheduled. The height of rectangles with red borders is the entities within the scheduled hours based on forecasted series, whereas the corresponding values after real-time scheduling based on forecast results are shown with green border rectangles. Because of accurate forecasts, in both cases (CO$_2$ intensity and electricity prices), the hours with minimum error values can be scheduled efficiently. However, it is worth noticing that there can be variations in scheduled hours based on the minimization of CO$_2$ intensity and electricity prices. As shown in Figure \ref{Fig2}, based on the forecasting results, for four hours scheduling on a specific day, the cheapest electricity can be obtained from 14th to 17th hours, whereas the least carbon-intensive hours are the 13th, 14th, 16th, and 17th. A slight variation in hours occurred for scheduled hours based on optimum (minimum) electricity prices and carbon intensities. Further, this variation goes on increasing with the increase in the number of hours to be scheduled. Since there is no strong correlation between these two parameters, it is obvious that the optimally scheduled hours based on one parameter will not be optimum for the other one. Hence, it becomes a crucial decision to find a trade-off between these two parameters and to schedule the hours such that inexpensive (not the cheapest) electricity can be purchased which will be responsible for lesser (not the least) carbon emissions. However, it will be important to observe at what cost the CO$_2$ emission in the electricity be reduced.

	\subsection{Trade-off between spot price and CO$_2$ intensity}
	
	This section proposes a graphical methodology to schedule the optimum hours based on the trade-off between electricity price and CO$_2$ intensity. The proposed methodology is based on simple geometrical principles. It uses forecasted values to schedule the optimum hours for the targeted day in the electricity market, and the values of previous days as a reference for the optimum scheduling.
	
\begin{figure}[ht]
		\centering
		\includegraphics[width=\textwidth]{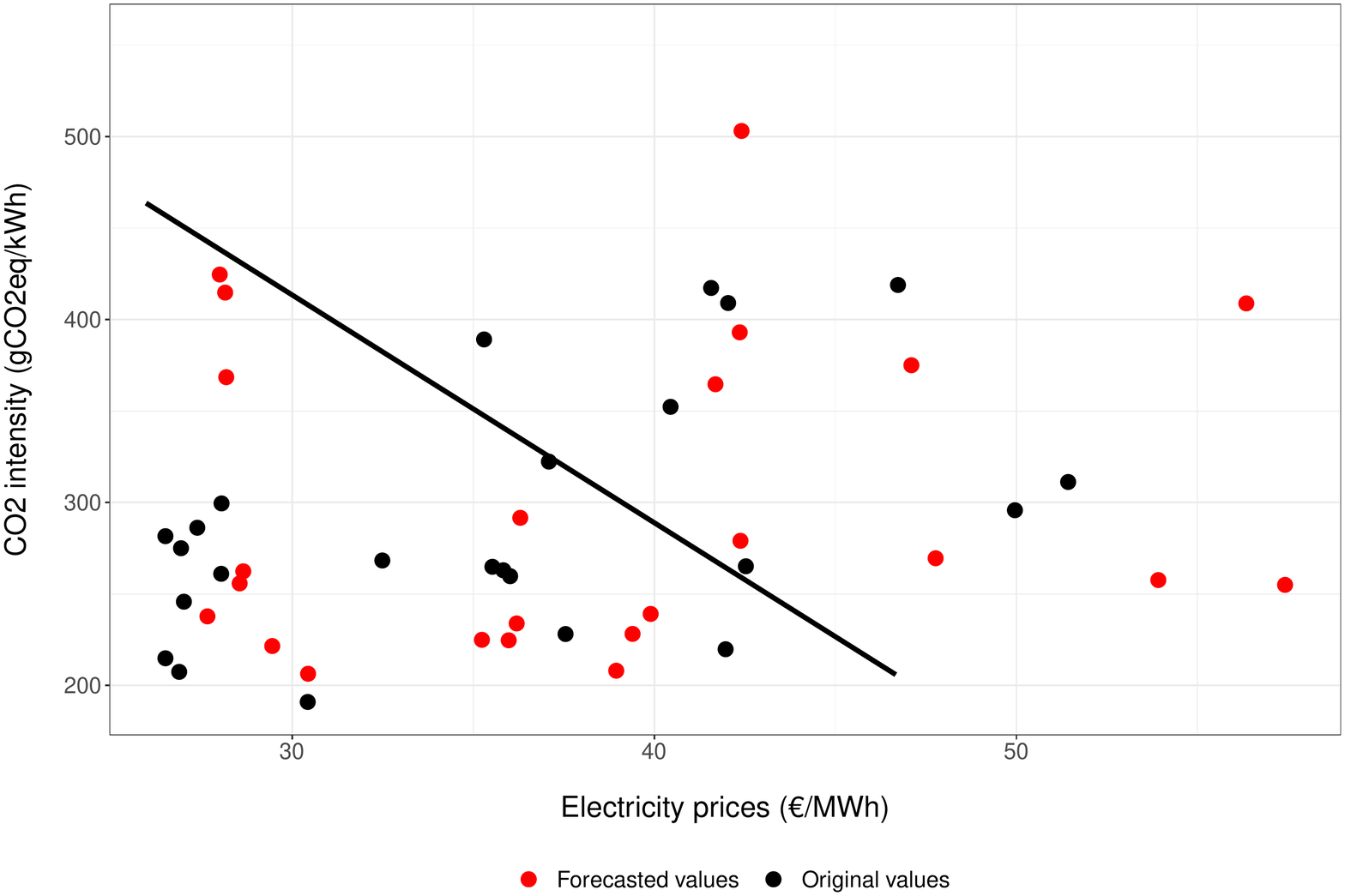}
		\caption{Scatter plot of hourly electricity price and CO$_2$ intensity. Historical values in black and forecasted values in red.}
		\label{line1}
	\end{figure}

	The first step of the methodology is the scatter plot of electricity prices (on the X-axis) versus CO$_2$ intensity (on the Y-axis) for the previous time horizon (day, month or year based on the targeted schedule). This is shown in Figure~\ref{line1}. The scatter plot shows the hourly relations between electricity prices and CO$_2$ intensity. The points in the scatter plot which are near to the origin of the plot are the hours with the lowest values of both entities (CO$_2$ intensity and electricity price). These are the preferred hours for optimum daily scheduling.
	
	The second step of the methodology is the selection of the desired number of optimum hours. This is done through plotting a line segment intersecting both the axes at a specific angle so that the desired number of optimum hours (points) can be collected between the origin of the plot and the line segment as shown in Figure \ref{line1}. The angle of the line represents the trade-off between the electricity price and CO$_2$ intensity.
	
	The line segment is a movable entity that starts from the origin of the plot and moved on the X-axis without changing its angle. The line segment stops moving when it has selected a specific number of points (representing hours) of the scatter plot between the origin of the plot and the line segment itself. These selected points (hours) are considered to be the reference for the optimum hours for scheduling.
	
\begin{figure}[ht]
		\centering
		\includegraphics[width=\textwidth]{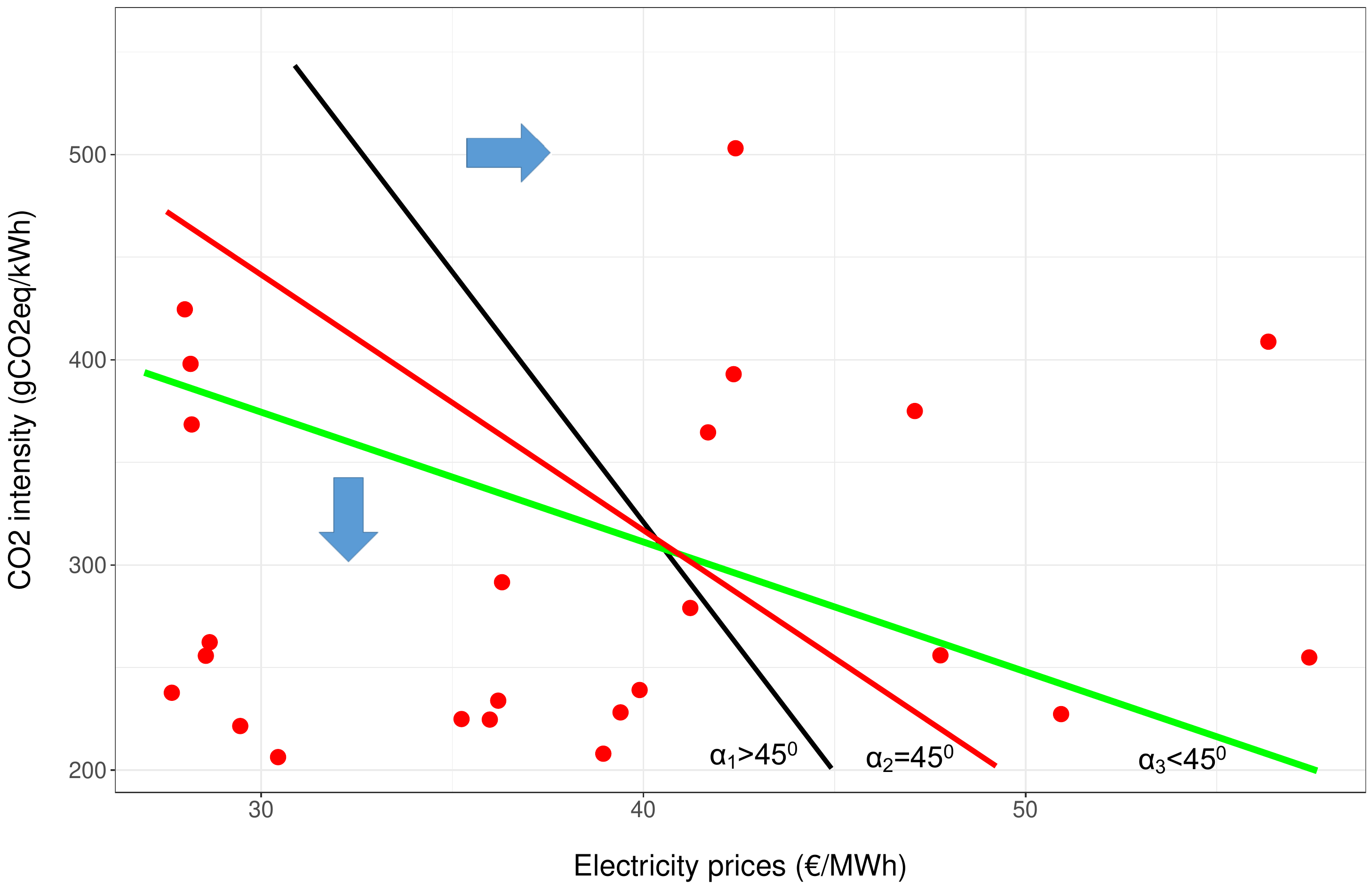}
		\caption{Visualizing different angles of the line used for scheduling. The steeper the line, the higher the weight assigned to electricity prices. The flatter the line, the higher the weight assigned to CO$_2$ intensity.}
		\label{rplot}
	\end{figure}

	The default angle of the line segment with the X-axis is 45 degrees, which ensures that the points (hours) selected by this line have an equal weight of both CO$_2$ intensity and electricity prices. This angle can be varied from 0 to 90 degrees as shown in Figure~\ref{rplot}. The minimum and maximum angles represent CO$_2$ intensity and electricity prices, respectively, whereas, the in-between angles represent the different ratios of the trade-off between CO$_2$ intensity and electricity prices. The line segments with an angle less than 45 degrees are more towards CO$_2$ intensity benefits and those with more than 45 degrees are towards better electricity prices. This is explored in Section~\ref{investigating-tradeoff}.
	
	The third step of the methodology is plotting the scatter plot of forecasted values for the targeted hours as shown in the red dots in Figure~\ref{line1}. The red dots lying below the line segments are considered as optimum hours for the given trade-off and are considered as suitable hours for the scheduling. It is possible that the number of hours scheduled based on the forecasted values (shown in red points) varies from that of the historic values. Hence, it is crucial to adjust this difference in the number of hours. 
	
\begin{figure}[p]
		\centering
		\begin{subfigure}{.7\textwidth}
			\centering
			\includegraphics[width=\textwidth]{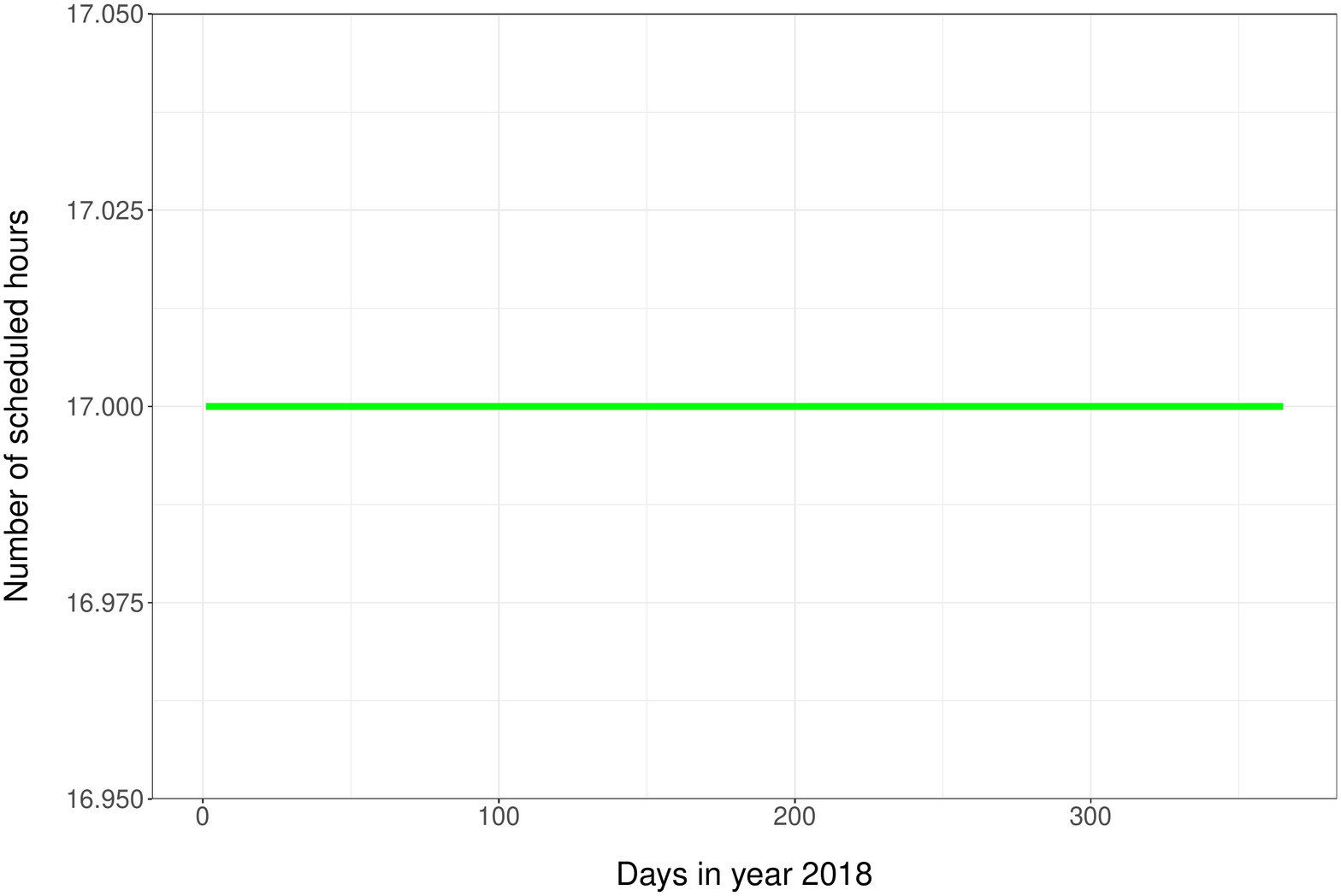}
			\caption{}
			\label{figxa}
		\end{subfigure}
		\begin{subfigure}{.7\textwidth}
			\centering
			\includegraphics[width=\textwidth]{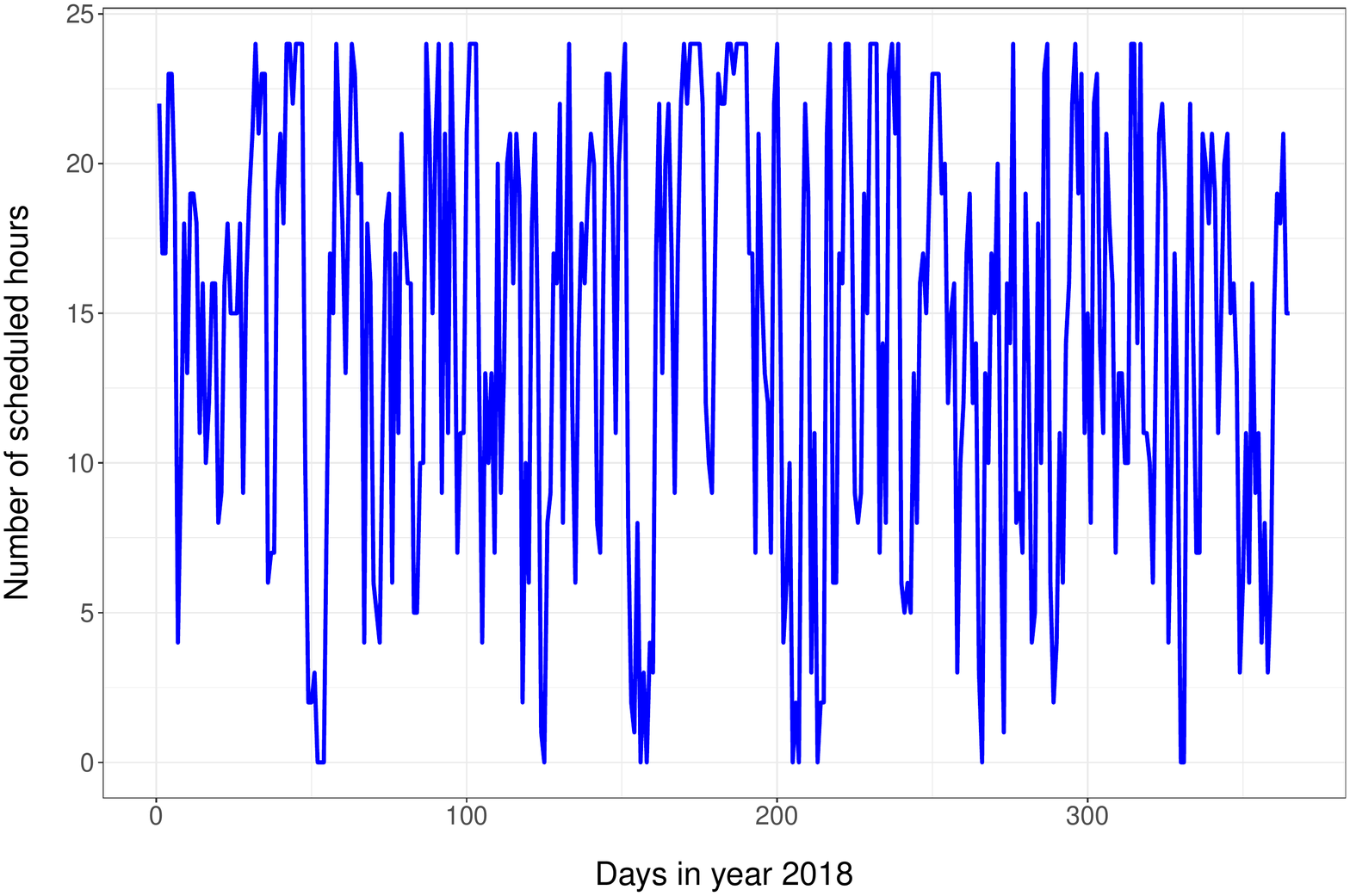}
			\caption{}
			\label{figxb}
		\end{subfigure}
		\begin{subfigure}{.7\textwidth}
			\centering
			\includegraphics[width=\textwidth]{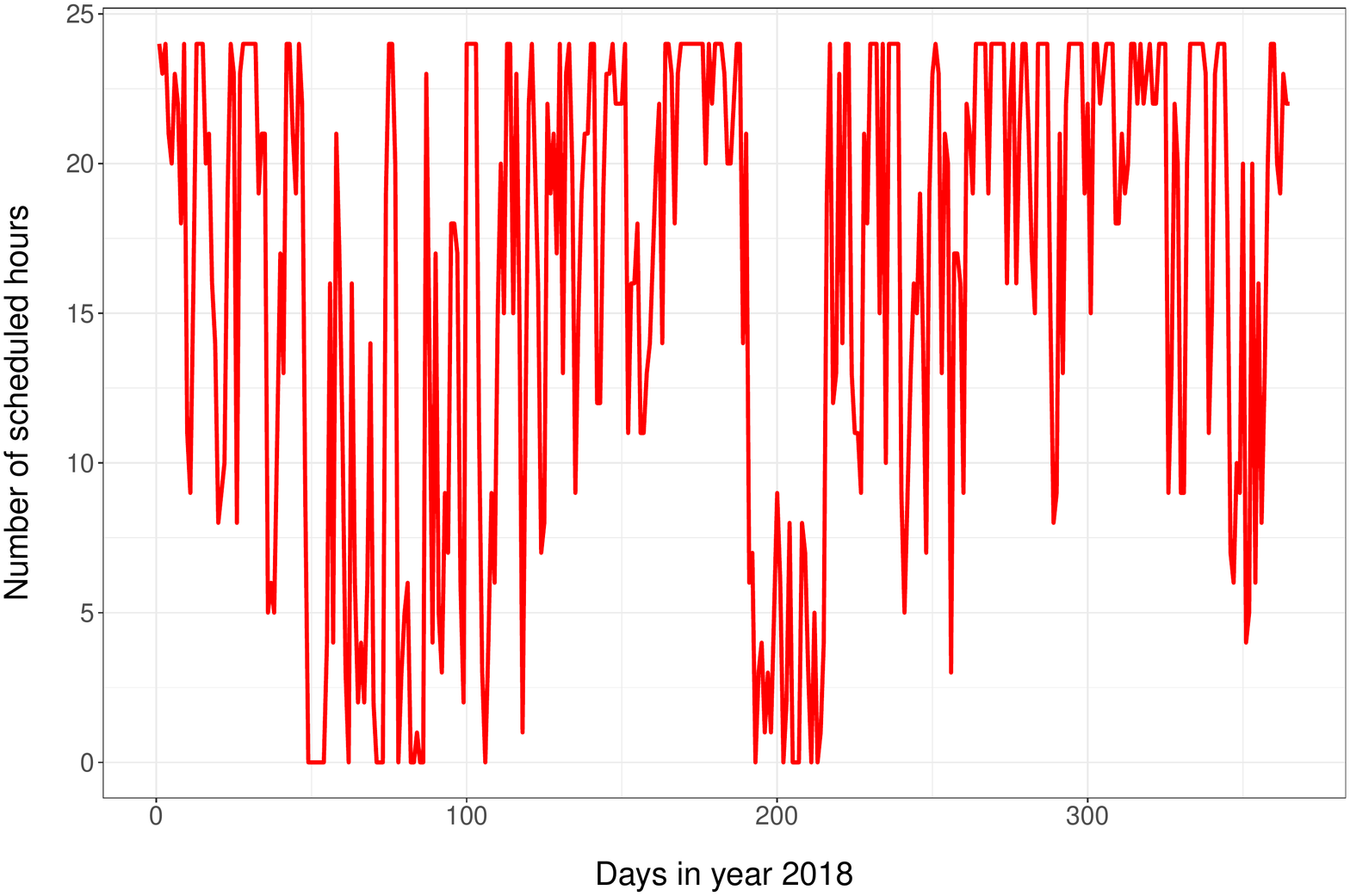}
			\caption{}
			\label{figxc}
		\end{subfigure}
		\caption{The number of scheduled hours per day of year 2018 in Denmark for (a) daily, (b) monthly, and (c) yearly scheduling.}
		\label{scheduling}
	\end{figure}

	In the proposed study, the hours are scheduled for a year with three strategies, i.e., daily, monthly, and yearly. For the daily strategy, the targeted number of FLHs is settled on a daily basis. For example, if the targeted FLHs is 12 hours and the number of hours scheduled with the proposed methods is 10. Then the remaining two hours are chosen from within the day, which will be from the set of hours above the 45-degree line (but, nearest to the line), whereas, in the monthly strategies, the difference of the number of hours scheduled based on historic and forecasted hours are adjusted at the end of each month. For instance, suppose 400 FLHs are to be scheduled per month and the number of hours scheduled for a month based on the proposed graphical method is 380. Then the remaining 20 hours are adjusted in the last days of that particular month. If, towards the end of the scheduling period, the number of remaining hours $X$ exceed, 24 the last $X$ hours of the scheduling period will be scheduled to meet the required amount of FLH. Similarly, for the yearly strategy, such remaining hours will be adjusted in the last days of the year. Based on these strategies, the number of hours scheduled per day is shown in Figure~\ref{scheduling}.
	
	For daily scheduling, the leftover hours are settled every day, therefore, targeted FLHs are get scheduled daily. Hence, the number of hours scheduled daily is uniform throughout the year as shown in Figure \ref{scheduling} (a). Whereas, in monthly and yearly schedules, the leftover hours are scheduled at the end of each month and end of the year, respectively; and these patterns can be observed in Figure \ref{scheduling} (b) and (c), respectively.
	
\begin{figure}[h]
		\centering{\includegraphics[width=0.8\textwidth]{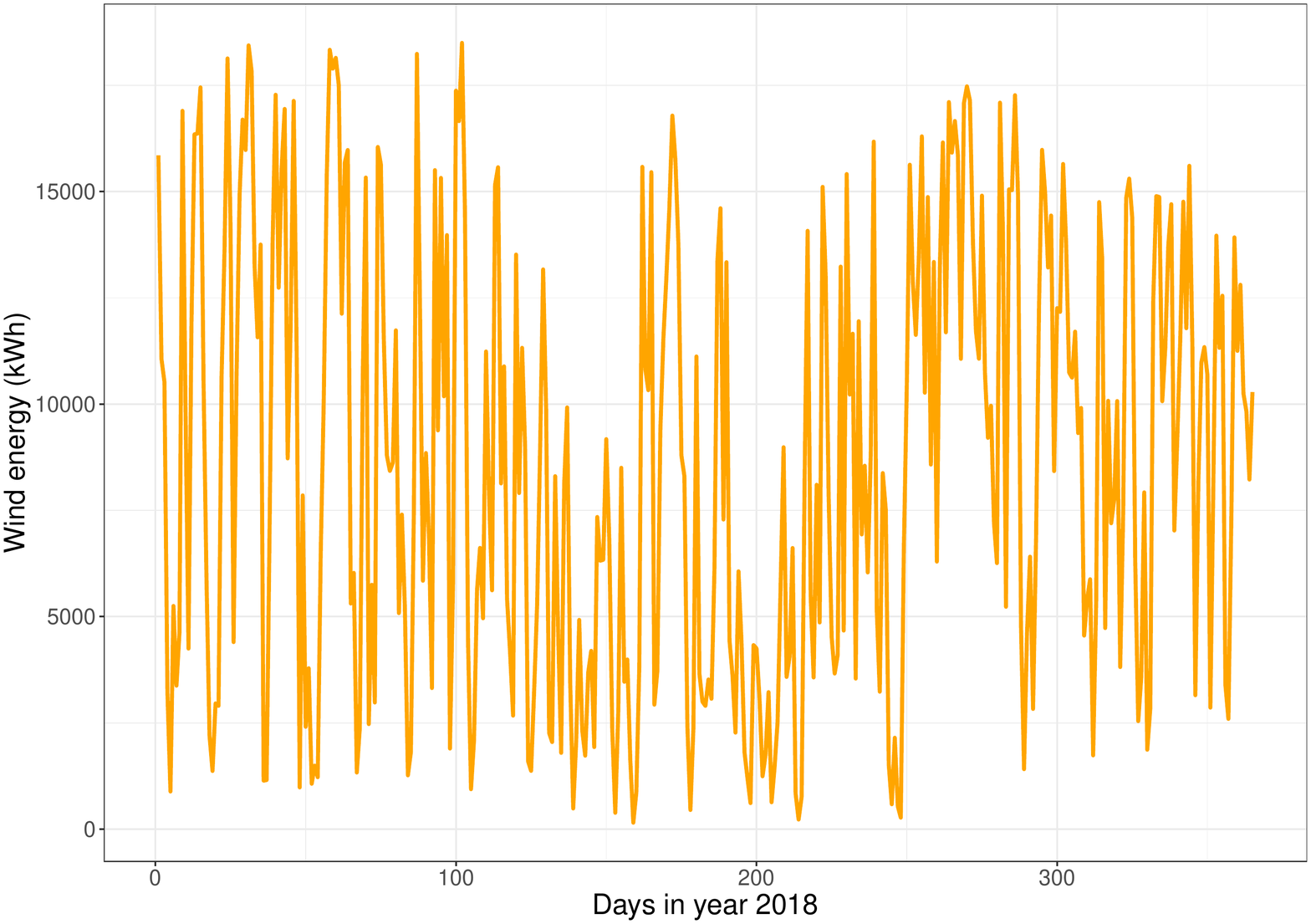}}
		\caption{Daily wind energy generated in Denmark for the year 2018.}
		\label{FigNew}
	\end{figure}

	It is worth noting how the scheduled hours are related to the share of the renewable energy produced in the respective days. In Figure~\ref{scheduling}, the hours scheduled per day are shown and it can be observed that the peaks in the plots are aligned with the patterns of daily wind energy generated in Denmark for the year 2018 as shown in Figure~\ref{FigNew}. For example, the minimum number of hours is scheduled in Figures~\ref{scheduling} (b) and (c) at around 200th day and this is analogous to the minimum amount of wind energy generated at the 200th day in Figure~\ref{FigNew}. Wind is the dominant renewable source of electricity in Denmark, thus this comparison between Figure~\ref{FigNew} and Figure~\ref{scheduling} (b) ensures the effectiveness of the proposed methodology in scheduling the hours with highest shares of renewable electricity production and minimum CO$_2$ emissions.
	
	\clearpage
	
	\section{Simulation study}
	
	\subsection{Simple optimization}
	
	Full Load Hours (FLH) is a way to characterize the power plant energy output over a period of time (usually a year). FLH states how many hours a plant or energy system have taken to produce a specific amount of energy if it has been operating at its full capacity.
	This section is focused on FLH scheduling based on the trade-off between CO$_2$ intensity and electricity price with the proposed graphical methodology. For each of these optimizations the weight between price and CO$_2$ is fixed to either best, worst or trade-off. A varying weight is explored in a later section.

\begin{figure}[p]
		\centering{\includegraphics[width=\textwidth]{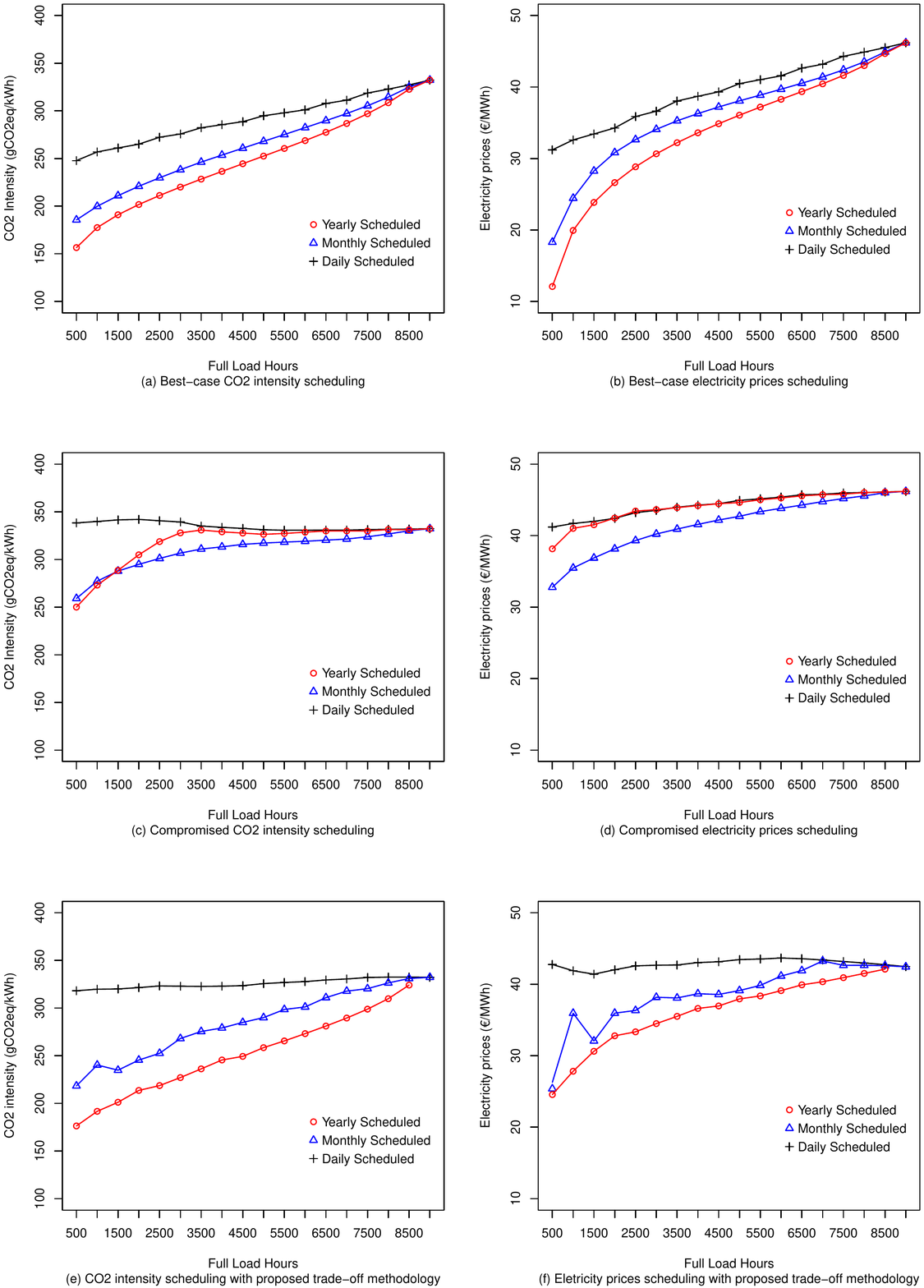}}
		\caption{Full load hours scheduling in Denmark.}
		\label{Fig4}
	\end{figure}

	Figure \ref{Fig4} (a) shows the per hour mean CO$_2$ emissions for given FLHs for the year 2018 in Denmark. These line-plots are plotted based on optimum (minimum) carbon-intensive hours scheduled with three different time-scales, i.e., optimum scheduling hours throughout the year (in red), averaged over the months (in blue) and averaged over the days (in black). In all cases the scheduling is carried out for an angle of 45 degrees for an equal weight of both CO$_2$ intensity and electricity price.
	This plot shows the amount of CO$_2$ emissions for the specific FLHs. Consider, it is of interest to know the amount of CO$_2$ emissions for 6000 FLHs in a power plant.
	Then for the yearly scheduling, optimum 6000 hours will be selected throughout the year and the mean of these hours will be the average CO$_2$ emissions at 6000 FLH. Whereas, for the monthly scheduling, (6000/12 months =) 500 optimum hours will be selected per month and the corresponding month will be the average CO$_2$ emissions for monthly scheduling at 6000 FLHs. Similarly, average CO$_2$ emissions  per day can be calculated by selecting (6000/(12 months X 30 days) = 16.67) $\sim 17$ optimum hours per day.
	
	From Figure \ref{Fig4} (a), it can be interpreted that on 6000 Full load hours, in the optimum case, on average, 268 gCO2eq/kWh CO$_2$ are emitted for the year 2018 in Denmark. These values are based on the original/historic CO$_2$ intensity time-series for the year 2018. Similarly, Figure \ref{Fig4} (b) shows a similar line-plot for per hour mean electricity prices for the respective FLHs.
	
	\newgeometry{textwidth=6in}
	\restoregeometry
	

	In both cases, the magnitudes of series are increasing with the increase in the FLHs. Throughout the year schedule, the rate of rising in the magnitude is lower as compared that of averaged over the monthly and daily intervals. This outcome is obvious, since the ample availability and choice of optimum hours throughout the year. On the contrary, for averaged daily scheduling, there are limited optimum hours per day.
	
	Again, it is worth observing the per hour mean electricity prices when best-case scheduling is done for the CO$_2$ intensity, and it is shown in Figure \ref{Fig4} (d). Similarly, the per hour mean carbon emission at the best case scheduling of electricity prices are shown in Figure \ref{Fig4} (c). In both figures (Figures \ref{Fig4} (c) and (d)), the per hour mean magnitudes are significantly higher than the best-case values at distinct FLHs. It means, for the scheduling with the best case of one parameter can be achieved with the compromise of the other one. Onward, such plots (shown in Figures \ref{Fig4} (c) and (d)) is referred as the `Compromised cases', throughout this article. Whereas, Figures \ref{Fig4} (e) and (f) are the CO$_2$ intensity and electricity price scheduling FLHs with the proposed trade-off methodology. Obviously, the amplitudes of both parameters are a bit higher than their respective best-case scheduling, but these are significantly lower than the amplitudes observed in the compromised cases shown Figures \ref{Fig4} (c) and (d), respectively. The observed improvements in CO$_2$ intensity and electricity price values achieved at the cost of a trade-off between both parameters and respective percentage improvements are tabulated for sample FLHs in Table \ref{T1}.

	Further, Figures \ref{Fig6}, \ref{Fig7}, and \ref{Fig8} shown in \ref{app} show the performance of the proposed trade-off methodology in Norway, France, and Germany, respectively. Though the trade-off scheduling observed at Denmark and Norway have shown significant improvements as compared to the compromised ones, improvements are negligible (or negative at some FLHS) for France and Germany. The quantified percentage improvements for 26 areas in Europe for yearly scheduling of 6000 FLHs are shown in Table \ref{T2}.

	\subsection{Effect of correlation between spot price and CO$_2$ intensity}
	
	As shown in Table \ref{T2}, for 26 areas, there are 2 to 41\% improvements in the CO$_2$ intensity values, whereas the improvements in 5 countries are negative. In terms of electricity prices, all countries have shown improvements varying in the range of 0.5 to 13\%. The negative values in the percentage improvements signify the degradation of CO$_2$ intensity scheduling due to the proposed trade-off methodology. Hence, it becomes important to understand the nature and relations of the electricity prices and CO$_2$ intensity patterns in various countries and to understand why some countries can not be beneficial with the trade-off methodology.
	
	\begin{figure}[h]
		\centering{\includegraphics[width=.9\textwidth]{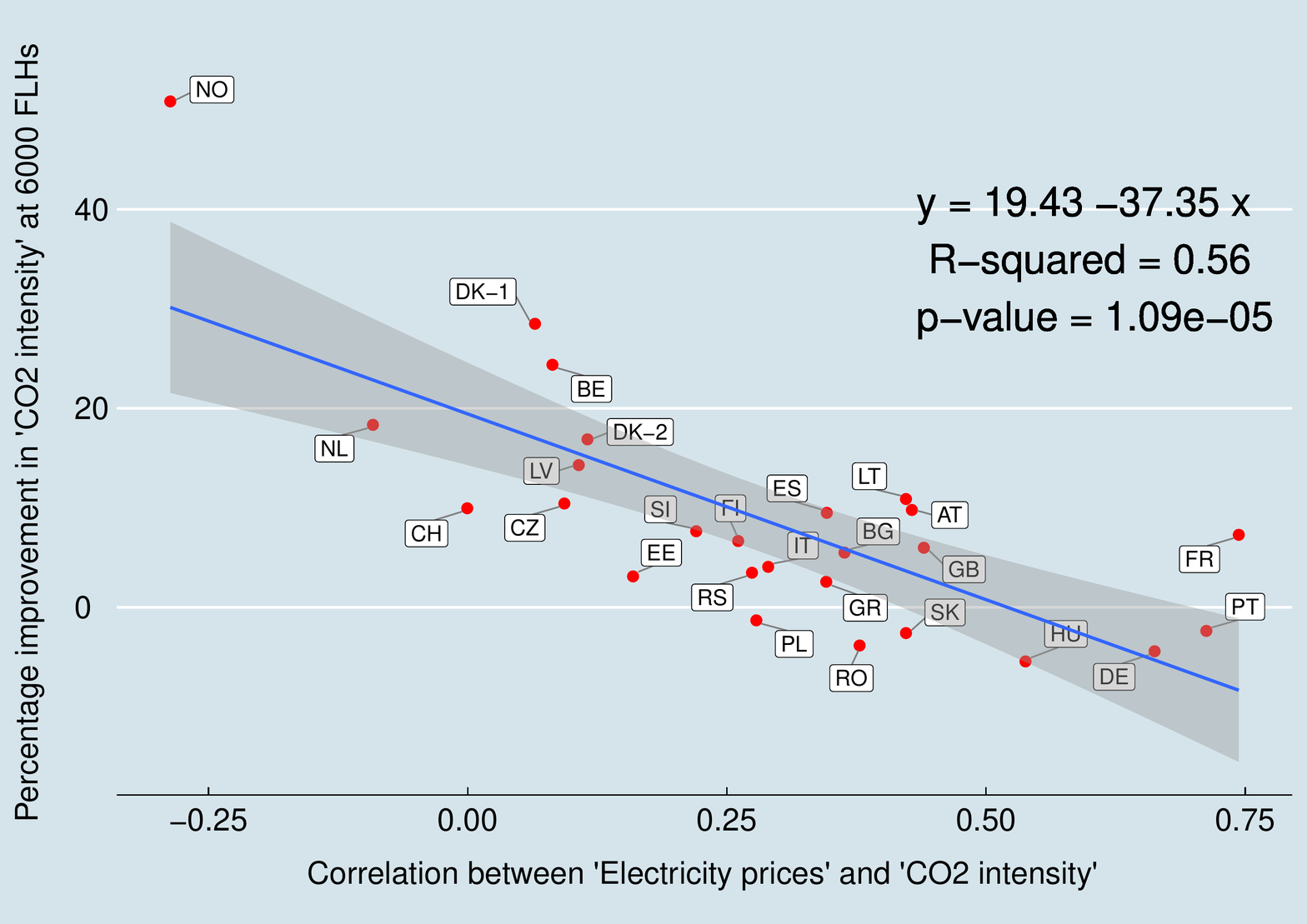}}
		\caption{Regression model showing the relation of `the correlation between electricity prices and CO$_2$ intensity' with `the percentage of improvements in CO$_2$ intensity with the proposed trade-off methodology at 6000 FLHs for yearly scheduling' for 26 areas for the year 2018.}
		\label{Fig9}
	\end{figure}
	
	Figure \ref{Fig9} shows the relation of the correlation between electricity prices and CO$_2$ intensity with the percentage of improvements in CO$_2$ intensity with the proposed trade-off methodology at 6000 FLHs for yearly scheduling for 26 areas. These parameters are successfully fitted into a linear regression model as shown in Figure \ref{Fig9} with significant p-value ($<< 0.005$). Also, the coefficient of regression ($R^2$) value is 0.5, which confirms the accuracy of the fitted regression model. This regression line confirms the indirect relationship between `the strong correlation in electricity prices and CO$_2$ emission' and `the percentage improvements in CO$_2$ intensity'. The countries such as FR, PT, DE, etc. which shows the strong correlation between electricity prices and CO$_2$ intensity values have shown negligible or negative improvements in the scheduling with the proposed trade-off methodology. On the contrary, the countries such as NO, DK, NL, BE, LV, etc, with a weak correlation of electricity prices and CO$_2$ intensity showed a significant and noticeable improvement in the scheduling.
	
\begin{figure}[p]
		\centering{\includegraphics[width=.8\textwidth]{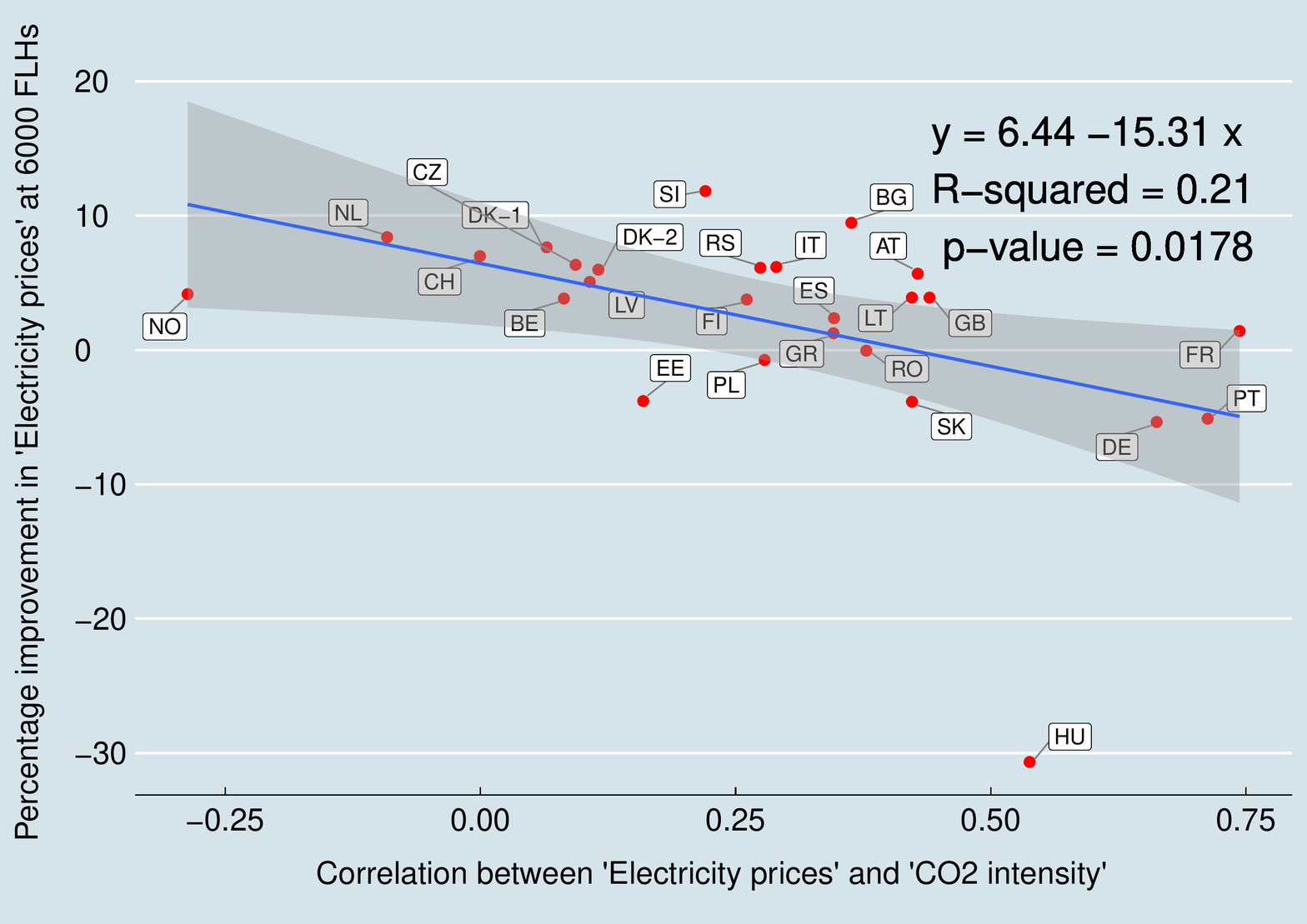}}
		\caption{Regression model showing the relation of `the correlation between electricity prices and CO$_2$ intensity' with `the percentage of improvements in electricity price with the proposed trade-off methodology at 6000 FLHs for yearly scheduling' for 26 areas for the year 2018.}
		\label{Fig10}
	\end{figure}

	Similarly, it is worth investigating the effect of the correlation coefficient on the improvements in the electricity prices. Figure \ref{Fig10} shows the fitted regression model among these parameters. The fitted model is weak (with insignificant p-value and very small regression coefficient of 0.02) and hence does not provide any valuable information.
	
\begin{figure}[p]
		\centering{\includegraphics[width=.8\textwidth]{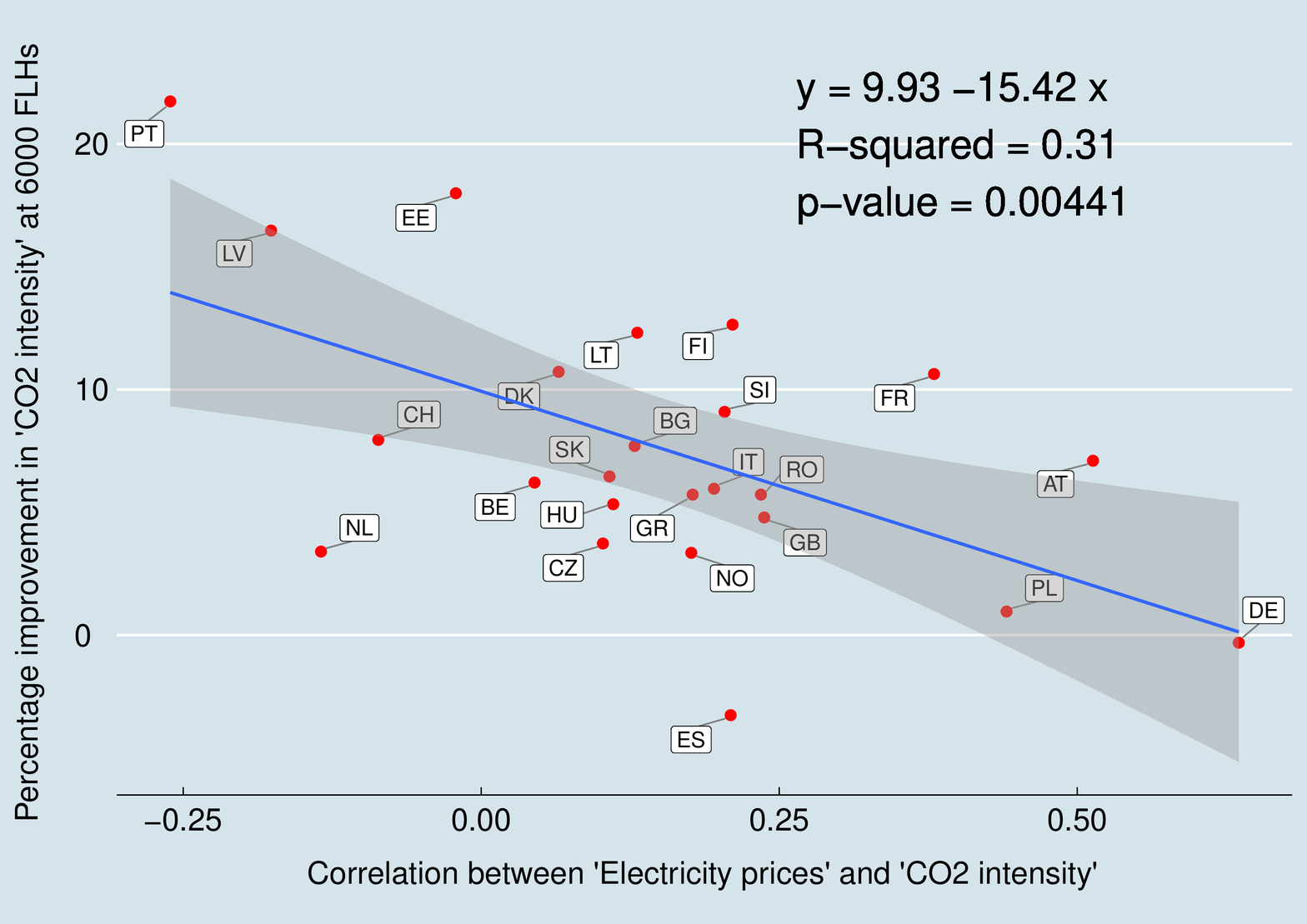}}
		\caption{Regression model showing the relation of `the correlation between electricity prices and CO$_2$ intensity' with `the percentage of improvements in CO$_2$ intensity with the proposed trade-off methodology at 6000 FLHs for yearly scheduling' for 26 areas for the year 2019.}
		\label{Fig11}
	\end{figure}

	Similar improvements are obtained for the dataset from the year 2019. Figure \ref{Fig11} shows the linear regression model for `percentage improvements in CO$_2$ intensity' versus the `correlation between electricity prices and CO$_2$ intensity' for the year 2019 for the same 26 European countries. The fitted model is not very strong (with regression coefficient ($R^2$) = 0.31), but acceptable to understand the relation between the targeted parameters for various countries. Similar to Figure \ref{Fig9}, Figure \ref{Fig11} shows the identical relationship for the year 2019. Most of the countries have shown consistent positions in the years 2018 and 2019 (as shown in Figures \ref{Fig9} and \ref{Fig11}). In a few countries (for example, PT), the CO$_2$ intensity and electricity prices have shown a weaker correlation in 2019 than in 2018, and therefore higher percentage improvements are achieved in the respective countries. These observations confirm the strong relations between `the correlation coefficient in electricity prices and CO$_2$ intensity' and `the percentage improvement in CO$_2$ intensity with the proposed methodology'.
	
	\clearpage	
	
	\section{Forecast study}
	
\begin{figure}[]
		\centering
		\begin{subfigure}{.9\textwidth}
			\centering
			\includegraphics[width=\textwidth]{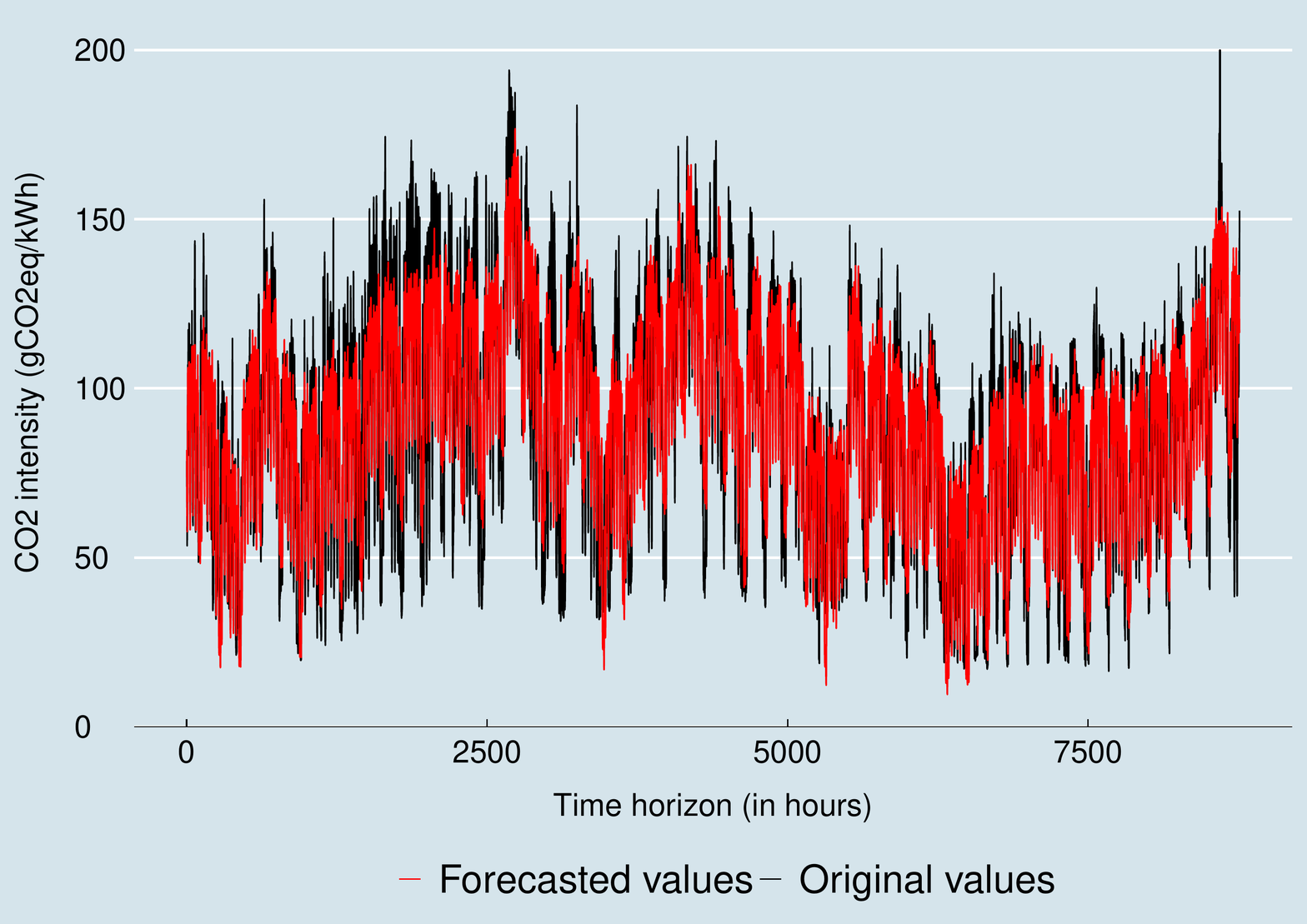}
			\caption{CO$_2$ intensity time-series.}
			\label{fig12a}
		\end{subfigure}
		\begin{subfigure}{.9\textwidth}
			\centering
			\includegraphics[width=\textwidth]{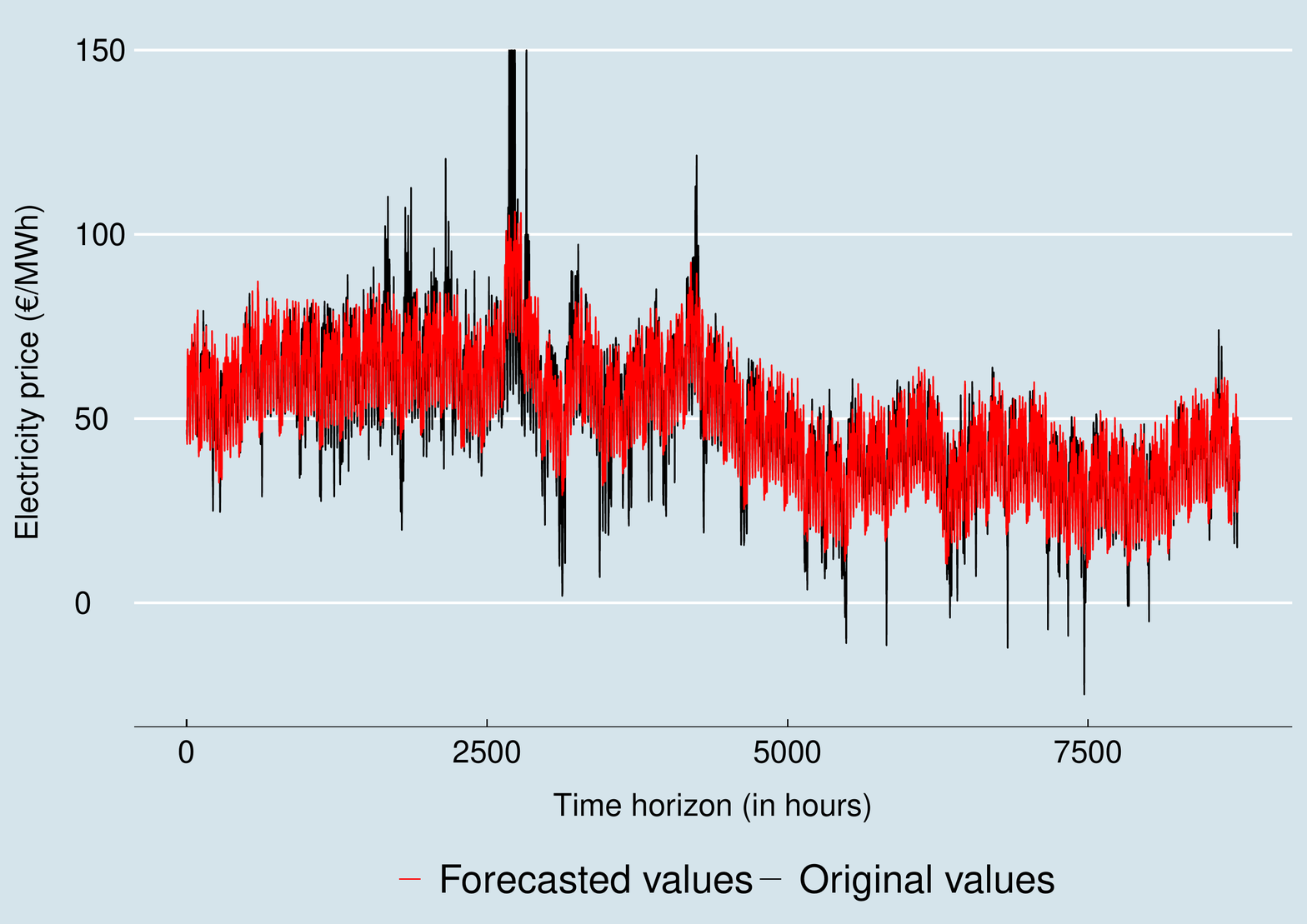}
			\caption{Electricity price time-series.}
			\label{fig12b}
		\end{subfigure}
		\caption{Historic and forecasted time-series of (a) CO$_2$ intensity and (b) electricity price for the targeted 365 days (from $2018-08-01\, 00:00:00$ to $2019-07-31\, 23:00:00$ at hourly interval).}
		\label{Fig12}
	\end{figure}

	Further, it is crucial to examine the performance of the proposed methodology on forecasted electricity prices and CO$_2$ intensity. For this purpose, the French electricity prices and CO$_2$ intensity time series for three months dated from $2018-05-01\, 00:00:00$ to $2018-07-31\, 23:00:00$ are used for training of the `Model 1' (which is discussed in the earlier sections) and forecasted 36 hours ahead values throughout the next year, from $2018-08-01\, 00:00:00$ to $2019-07-31\, 23:00:00$ with an iterated approach that was performed with the ForecastTB tool \cite{bokde2020forecasttb}. Out of these 36 hours, 13th to 36th hours (last 24 values) are considered as a part of hours to be used for the scheduling and the initial 12 hours are skipped due to the nature and rules of the European electricity spot markets. Figures \ref{fig12a} and \ref{fig12b} shows the historic and forecasted time-series of CO$_2$ intensity and electricity prices for the targeted 365 days (from $2018-08-01\, 00:00:00$ to $2019-07-31\, 23:00:00$), respectively. The statistical characteristics of both parameters in historic and forecasted time series are shown in Table \ref{charac} along with the correlation coefficients between them.
	
\begin{table}[h]
	\centering
	\caption{Statistical characteristics of historic and forecasted CO$_2$ intensity and electricity price time series. (Min. = Minimum, 1st Qu. = First quartile, Max. = Maximum, SD = Standard Deviation, CC = Correlation Coefficient.)}
	\label{charac}
	\begin{tabular}{ccccccccc}
		\hline
		Time series                                                             & Min.   & 1st Qu. & Median & Mean  & 3rd Qu. & Max.   & SD    & CC                    \\ \hline
		\begin{tabular}[c]{@{}c@{}}Historic CO$_2$\\ intensity\end{tabular}        & 16.35  & 61.03   & 88.78  & 87.26 & 111.65  & 200.00 & 33.99 & \multirow{2}{*}{0.81} \\
		\begin{tabular}[c]{@{}c@{}}Forecasted CO$_2$\\ intensity\end{tabular}      & 9.47   & 70.10   & 90.41  & 89.30 & 109.64  & 176.78 & 28.30 &                       \\ \hline
		\begin{tabular}[c]{@{}c@{}}Historic \\electricity prices\end{tabular}   & -24.92 & 36.74   & 47.95  & 49.37 & 62.60   & 150.00 & 18.27 & \multirow{2}{*}{0.87} \\
		\begin{tabular}[c]{@{}c@{}}Forecasted\\ electricity prices\end{tabular} & 9.50   & 38.17   & 50.37  & 50.52 & 63.68   & 106.23 & 16.93 &                       \\ \hline
	\end{tabular}
\end{table}
	
\begin{figure}[]
		\centering
		\begin{subfigure}{.9\textwidth}
			\centering
			\includegraphics[width=\textwidth]{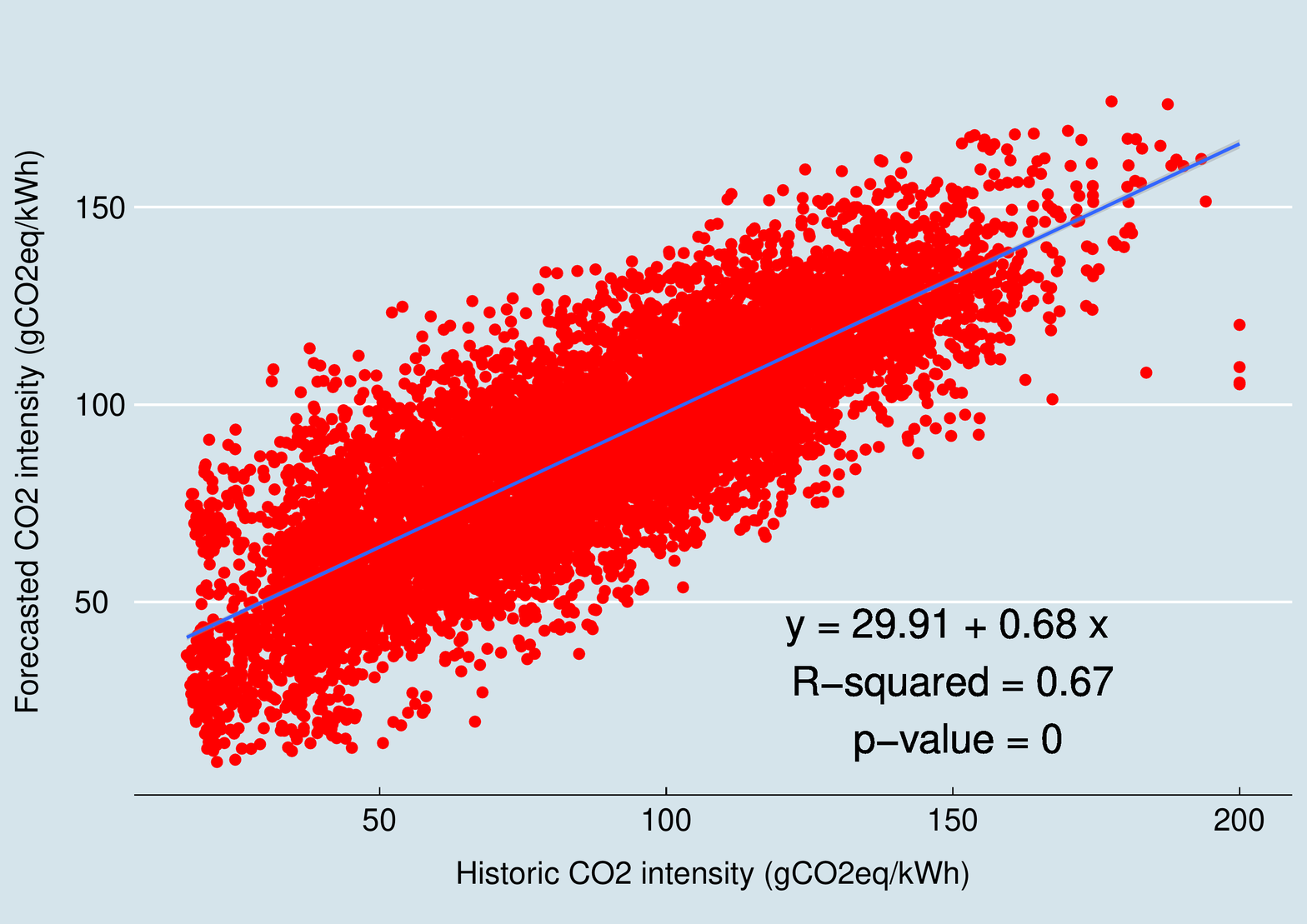}
			\caption{Linear regression model fitted for CO$_2$ intensity.}
			\label{fig13a}
		\end{subfigure}
		\begin{subfigure}{.9\textwidth}
			\centering
			\includegraphics[width=\textwidth]{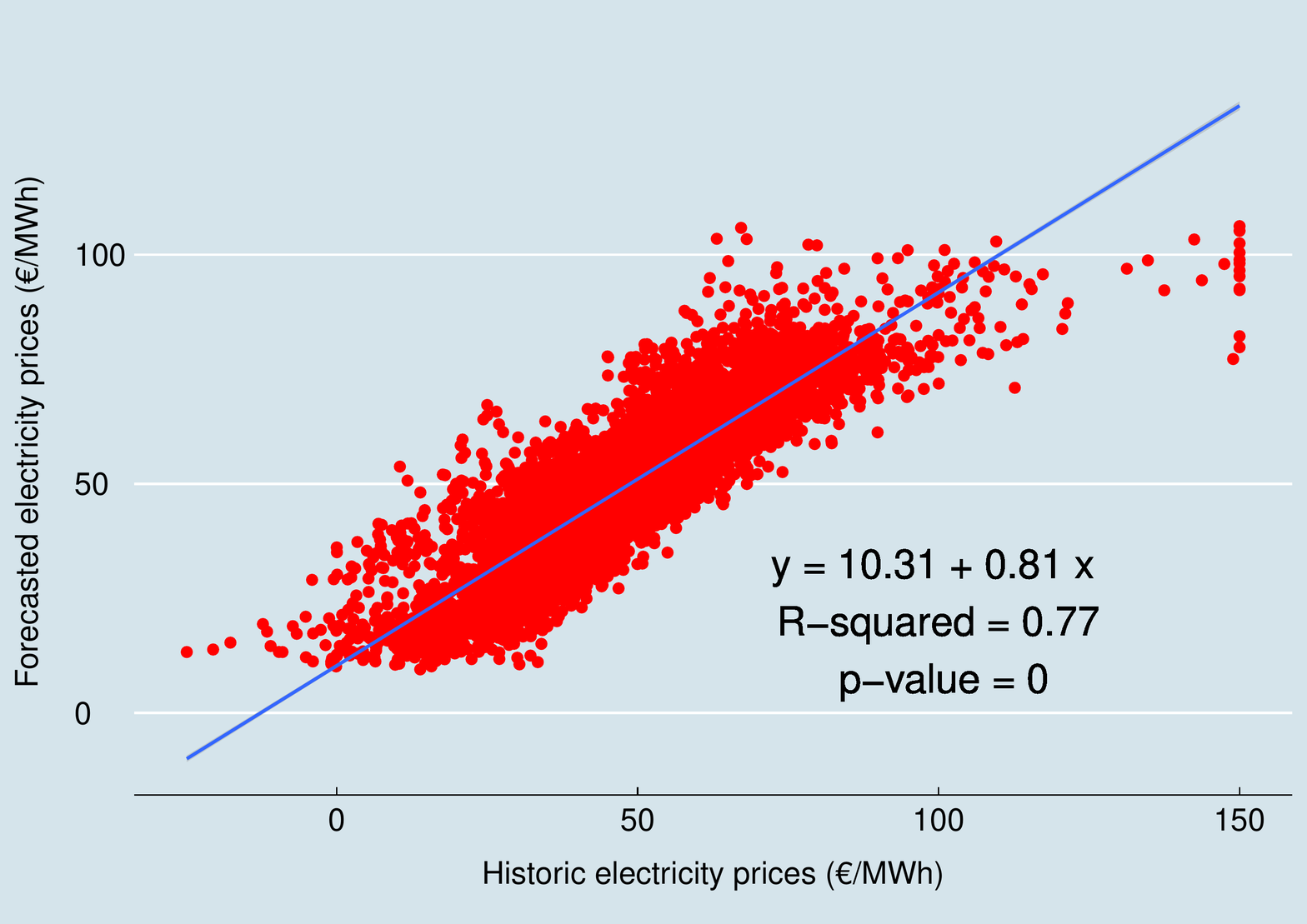}
			\caption{Linear regression model fitted for  electricity price.}
			\label{fig13b}
		\end{subfigure}
		\caption{Linear regression model fitted for forecasted and historic (a) CO$_2$ intensity and (b) electricity prices.}
		\label{Fig13}
	\end{figure}

	The forecasted results are validated with a linear regression model fitted in Figures \ref{fig13a} and \ref{fig13b} for CO$_2$ intensity and electricity prices, respectively. Both fitted models are fitted accurately with strong regression coefficients ($R^2$) of 0.67 and 0.77 for CO$_2$ intensity and electricity prices, respectively.
	
\begin{figure}[p]
		\begin{subfigure}{\textwidth}
			\centering
			\includegraphics[width=\textwidth]{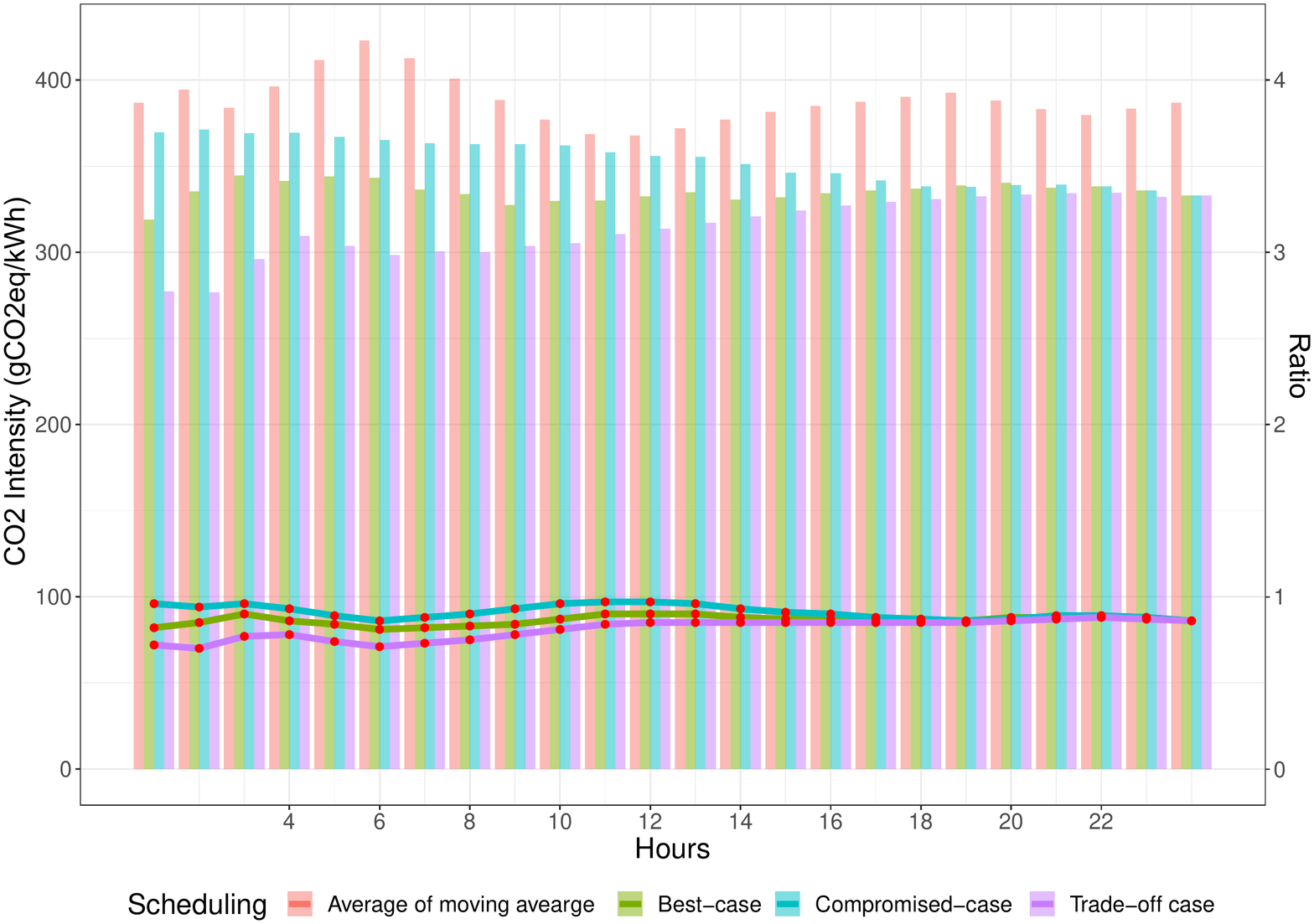}
			\caption{}
			\label{Fig14a}
		\end{subfigure} \\
		\begin{subfigure}{\textwidth}
			\centering
			\includegraphics[width=\textwidth]{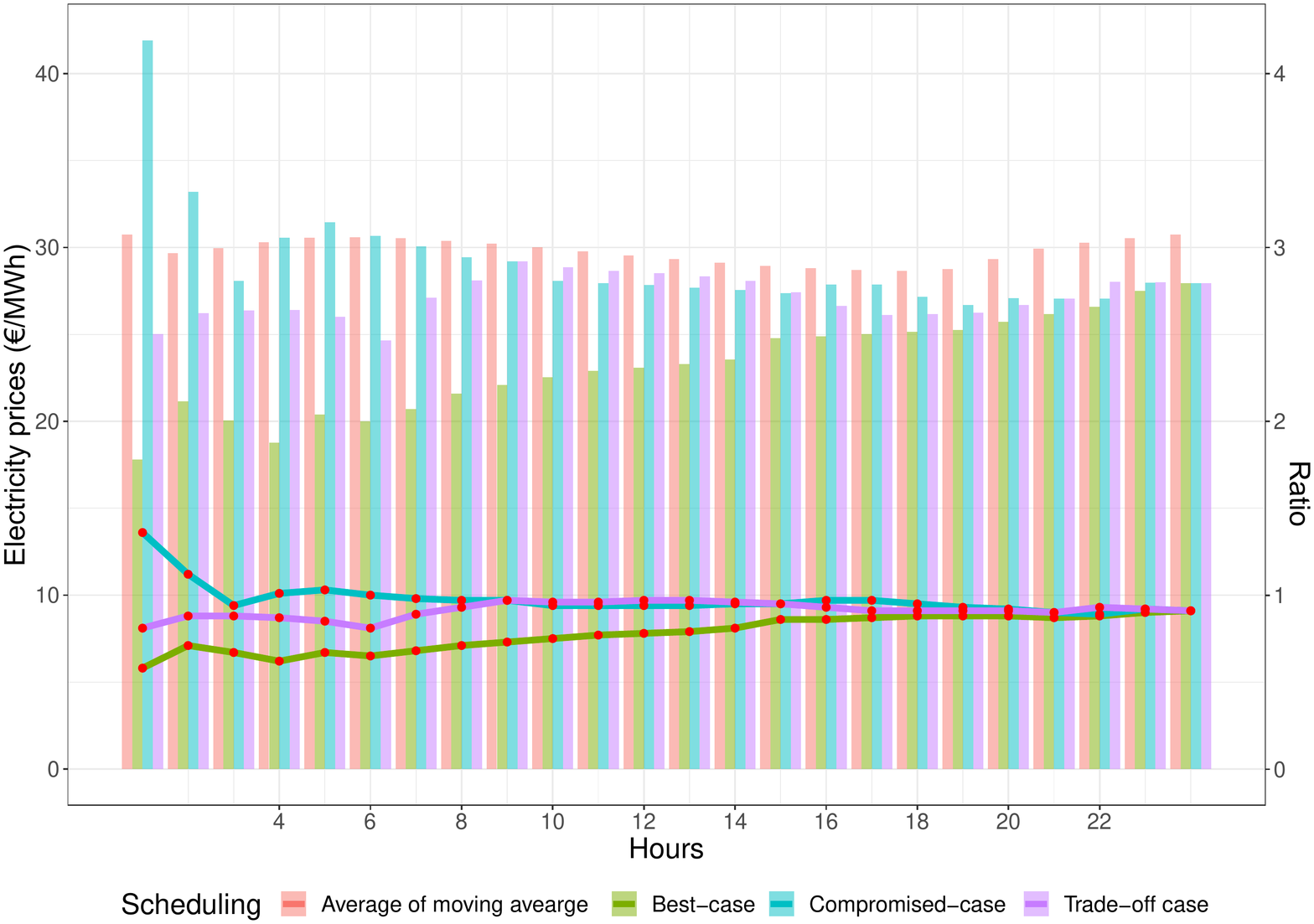}
			\caption{}
			\label{Fig14b}
		\end{subfigure}
		\caption{Comparison of emissions between (a) CO$_2$ intensity and (b) electricity price scheduling and consuming at a random time
			during the 24-hour interval for an entire year in Denmark. Each bar shows results for different
			durations of flexible consumption. The ratio between the bars is shown as the red line.
		}
		\label{Fig14}
	\end{figure}

\begin{figure}[p]
		\begin{subfigure}{\textwidth}
			\centering
			\includegraphics[width=\textwidth]{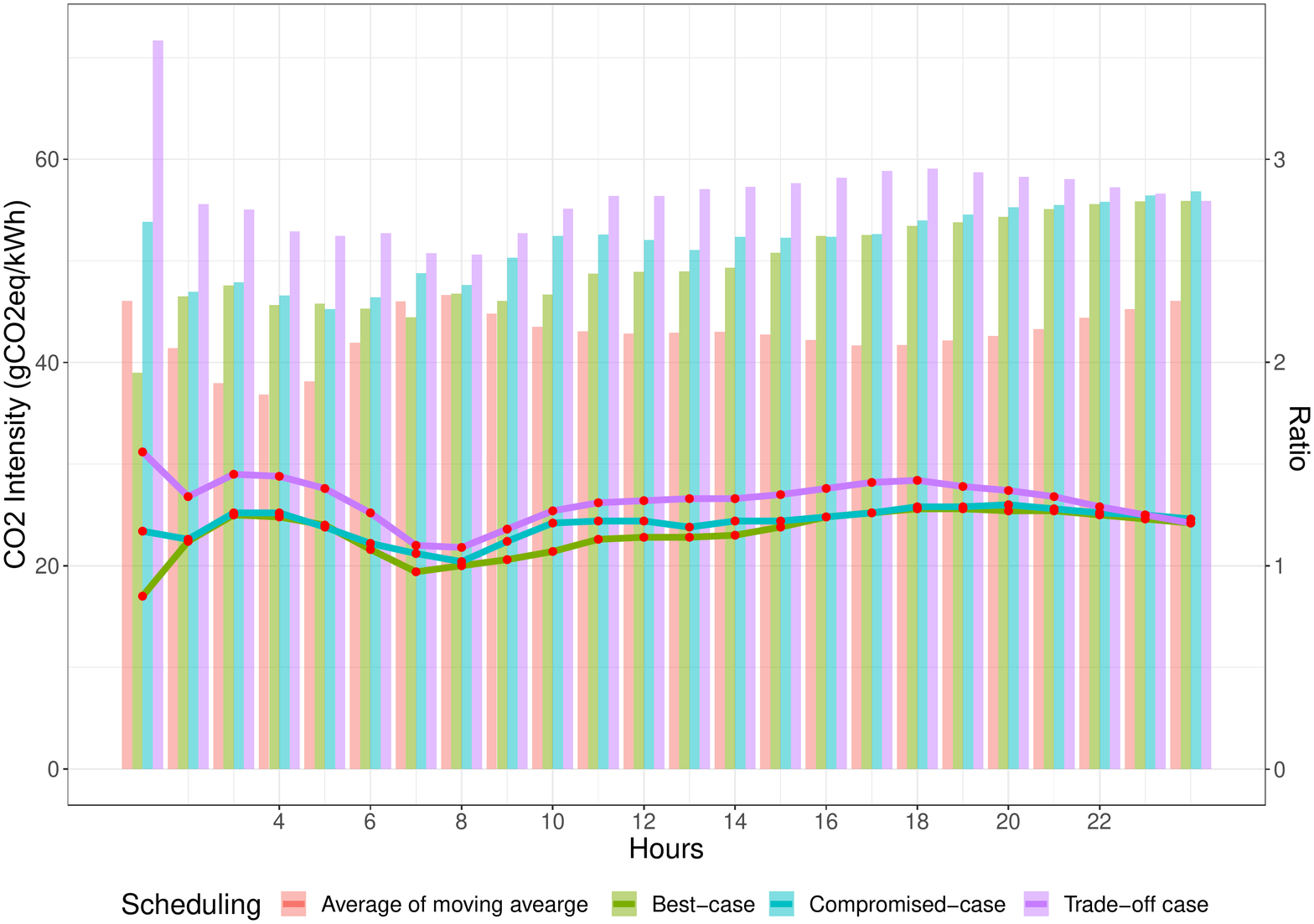}
			\caption{}
			\label{Fig15a}
		\end{subfigure} \\
		\begin{subfigure}{\textwidth}
			\centering
			\includegraphics[width=\textwidth]{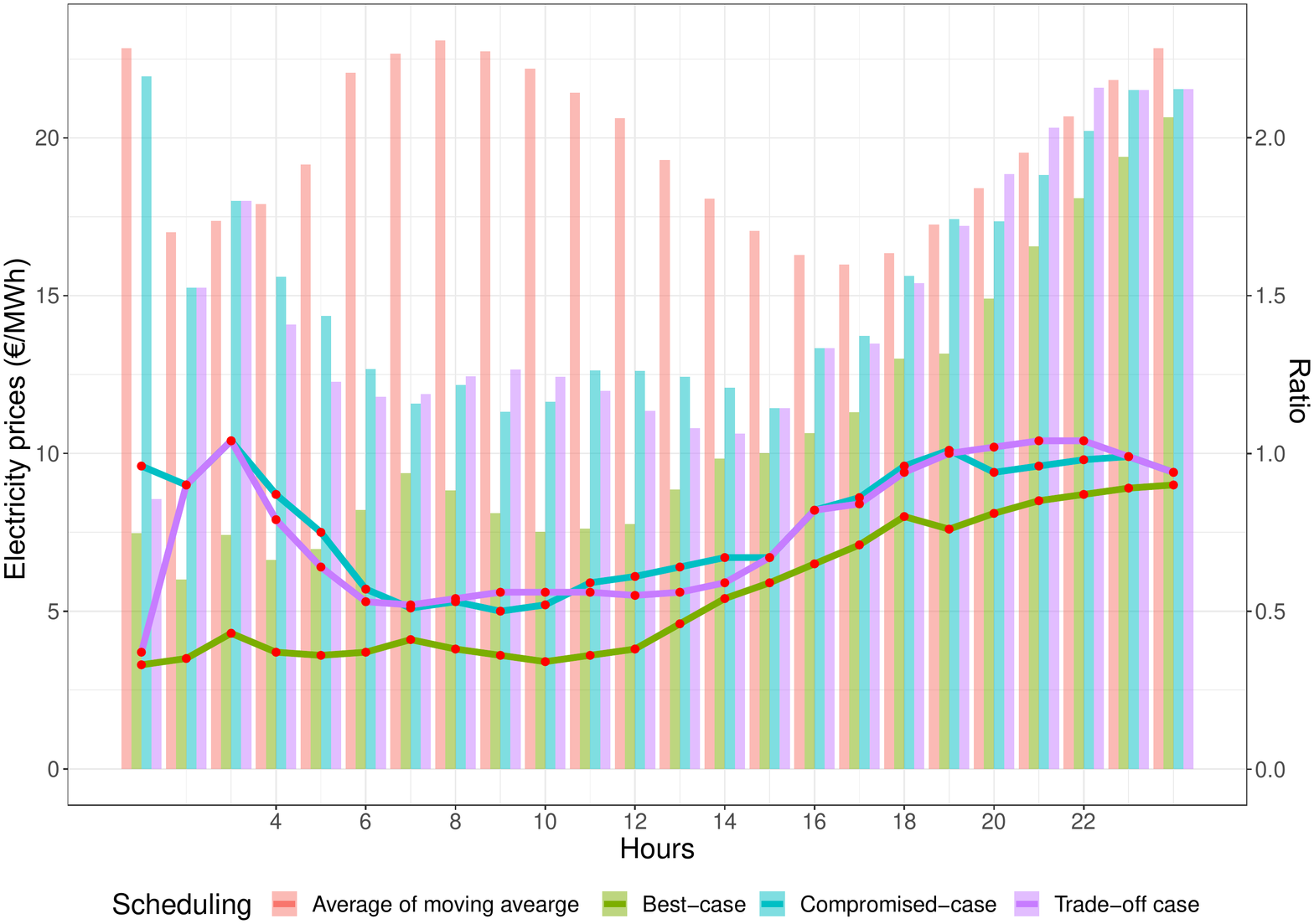}
			\caption{}
			\label{Fig15b}
		\end{subfigure}
		\caption{Comparison of emissions between (a) CO$_2$ intensity and (b) electricity price scheduling and consuming at a random time
			during the 24-hour interval for an entire year in France. Each bar shows results for different
			durations of flexible consumption. The ratio between the bars is shown as the red line.
		}
		\label{Fig15}
	\end{figure}

	Figure \ref{Fig14a} compares CO$_2$ intensity for consumption at a random time during the 24-hour interval for an entire year in Denmark and corresponding scheduled values at best, compromised, and trade-off cases of scheduling. Each bar shows results for different durations of flexible consumption from one to 24 hours. The ratio between the bars is shown as the respective colored lines. A ratio of one means there are no savings from scheduling compared to consuming at a random time. This is the case when the flexible consumption takes up all 24 hours of the window. The ratio decreases as the duration of flexible consumption decreases. This is to be expected as a shorter duration of consumption means more flexibility for scheduling during the 24 windows. Further, Figure \ref{Fig14b} compared the performance of scheduling cases for electricity prices during the 24-hour interval for an entire year in Denmark. A similar comparison is done for CO$_2$ intensity and electricity prices in France and is shown in Figure \ref{Fig15a} and \ref{Fig15b}, respectively.

	These comparison results confirm the negative relationship of `the correlation between electricity prices and CO$_2$ intensity' with `the percentage of improvements in CO$_2$ intensity with the proposed trade-off methodology' discussed in Figure \ref{Fig9}. In the case of Denmark, a significant improvement is observed for CO$_2$ intensity as well as electricity prices with the proposed graphical scheduling methodology. Whereas, the performance of the same proposed methodology observed similar or worse than the compromised cases for both CO$_2$ intensity and electricity price schedule.

	\clearpage
	
	\section{Investigating the trade-off between spot price and CO$_2$ intensity}\label{investigating-tradeoff}
\begin{figure}[ht]
		\centering
		\includegraphics[width=0.7\textwidth]{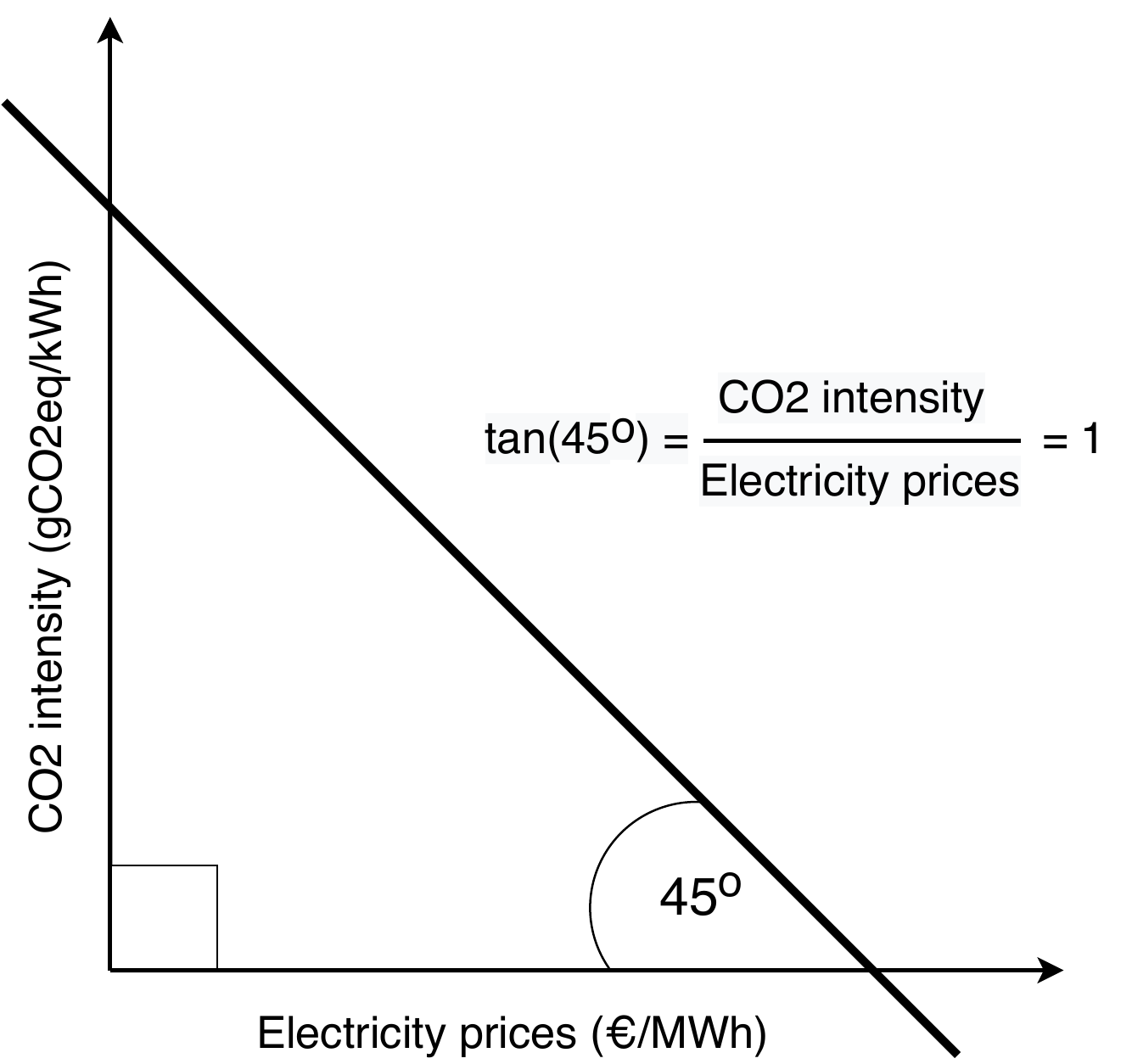}
		\caption{Schematic of equal weight between electricity price and CO$_2$ intensity.}
		\label{angle1}
	\end{figure}

	So far, we have discussed the scheduling of hours with the proposed graphical methodology with an inclined line at a 45 degree angle, which ensures an equal weight of the CO$_2$ intensity and electricity prices as shown in Figure~\ref{angle1}.
	
\begin{figure}[ht]
		\centering
		\includegraphics[width=0.7\textwidth]{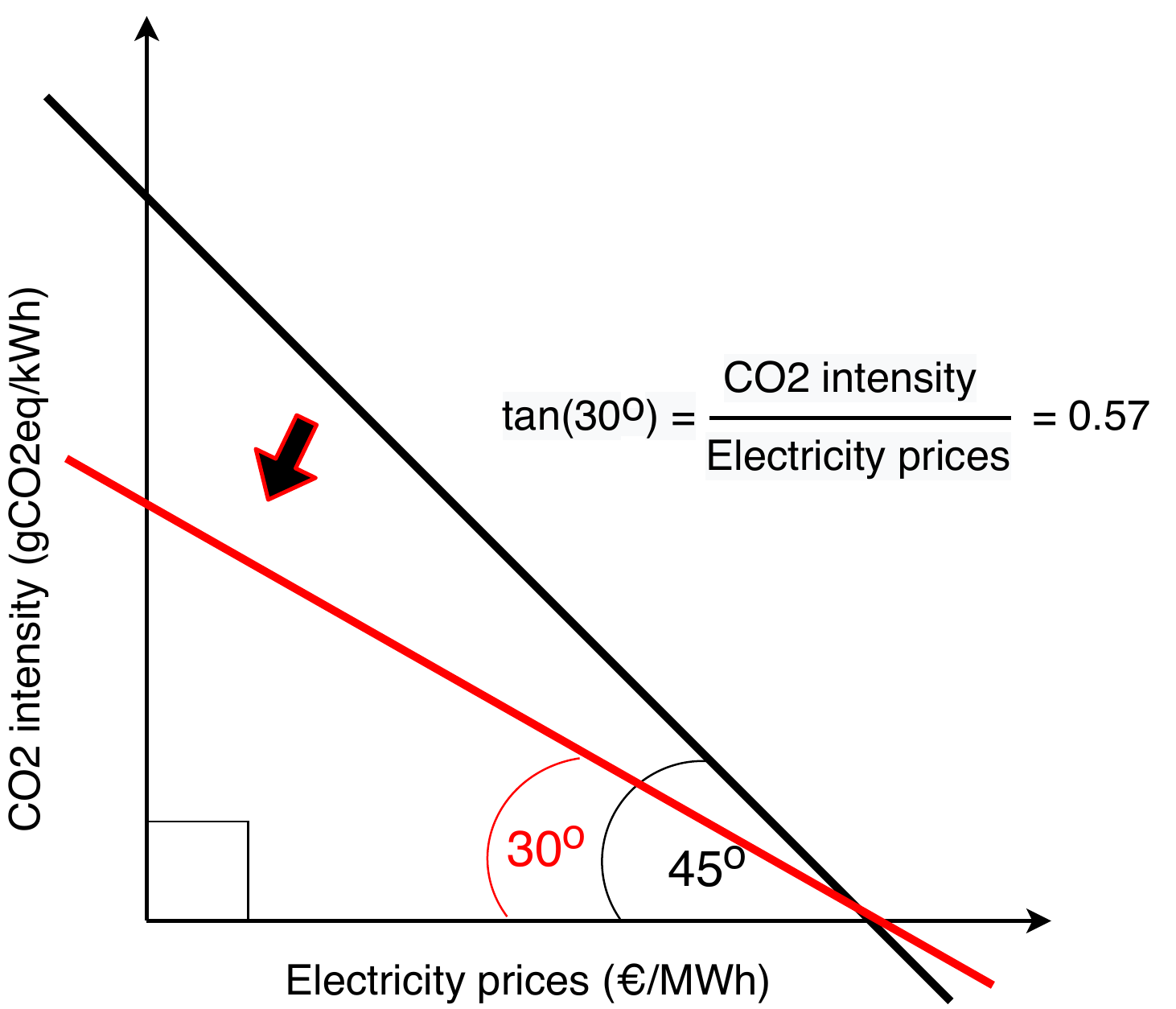}
		\caption{Schematic of impact on weight between electricity price and CO$_2$ intensity when changing the angle from 45 to 30 degrees.}
		\label{angle2}
	\end{figure}

	The proposed methodology allows us to change the share of CO$_2$ intensity and electricity prices by altering the angle of the inclined line. For instance, if we change the angle of inclination from 45 to 30 degrees, it leads to the change in the ratios of CO$_2$ intensity and electricity prices as shown in Figure~\ref{angle2}.
	
\begin{figure}[p]
		\centering
		\begin{subfigure}{.7\textwidth}
			\centering
			\includegraphics[width=\textwidth]{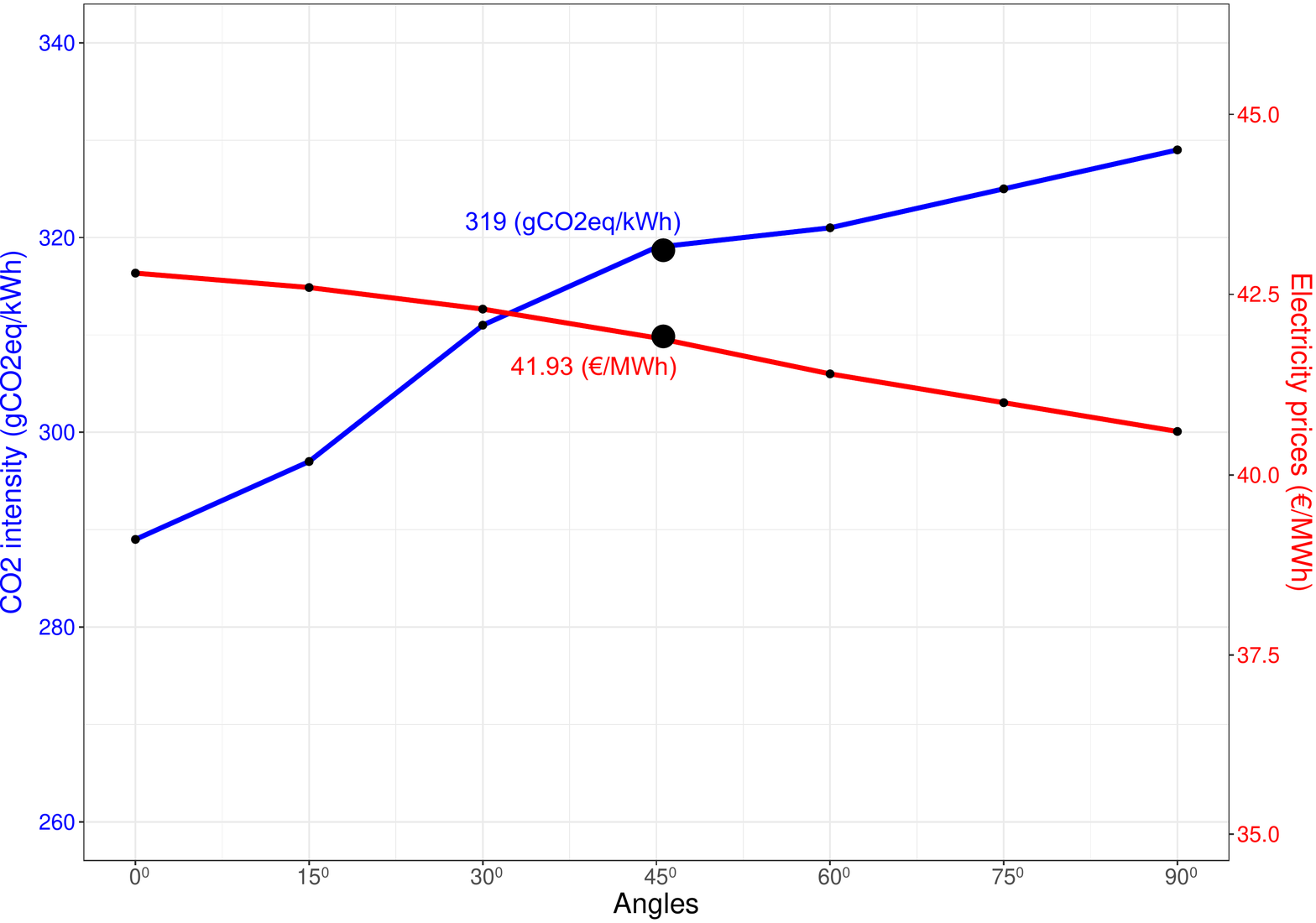}
			\caption{}
			\label{tradeoffa}
		\end{subfigure}
		\begin{subfigure}{.7\textwidth}
			\centering
			\includegraphics[width=\textwidth]{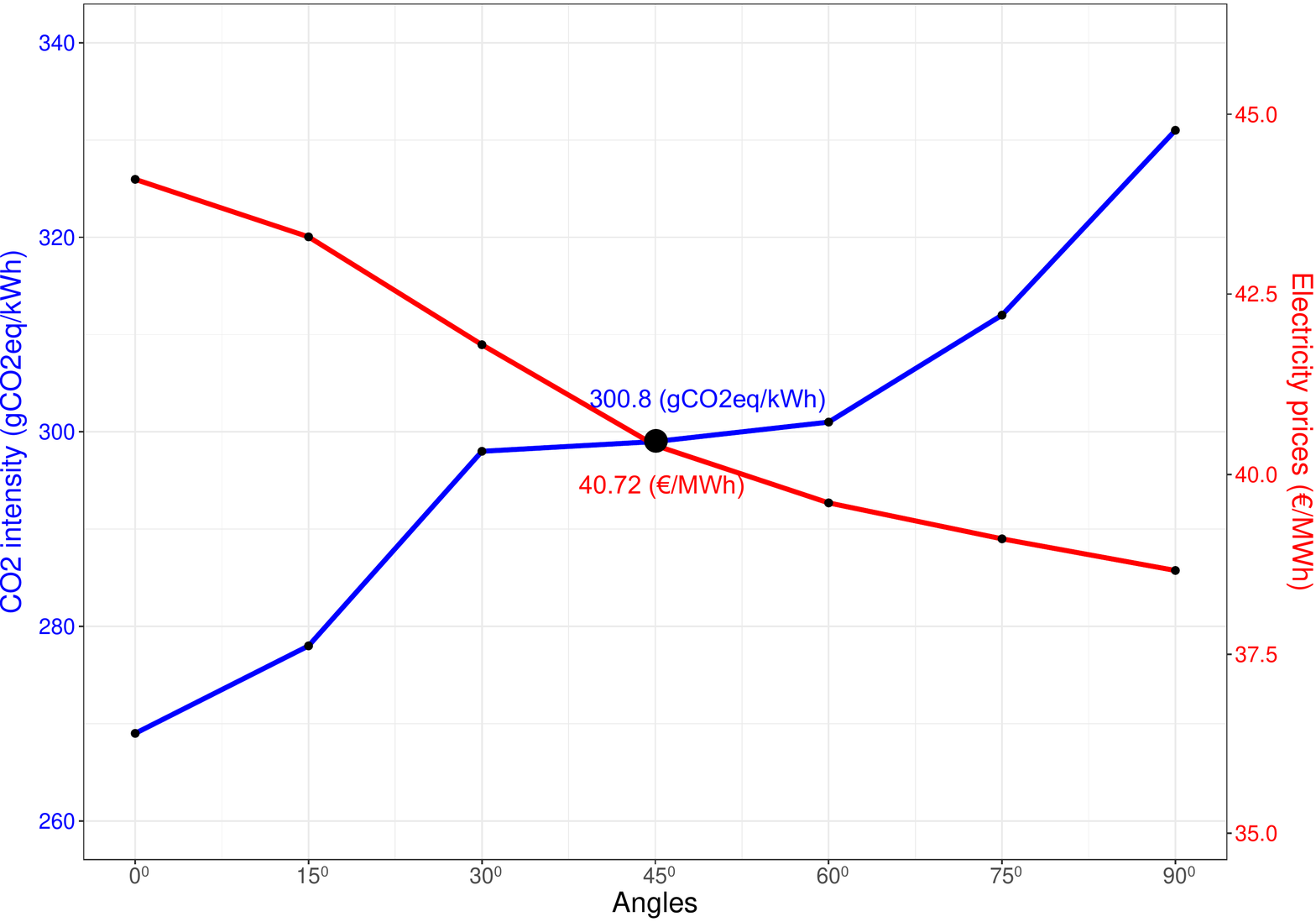}
			\caption{}
			\label{tradeoffb}
		\end{subfigure}
		\begin{subfigure}{.7\textwidth}
			\centering
			\includegraphics[width=\textwidth]{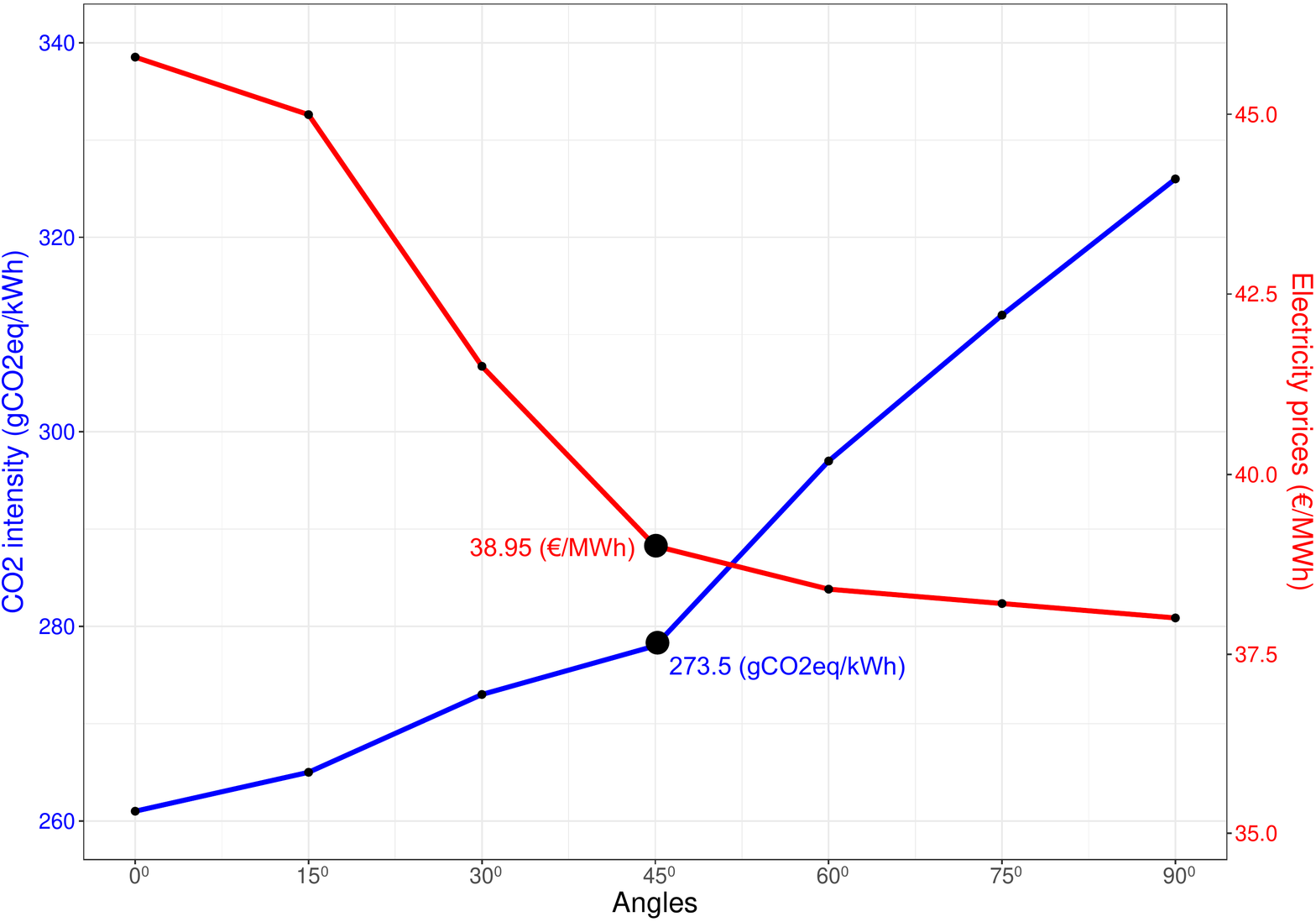}
			\caption{}
			\label{tradeoffc}
		\end{subfigure}
		\caption{Observed changes in the CO$_2$ intensity and electricity prices for 6000 FLHs with respect to change in the inclination angle of the line for a) daily, b) monthly and c) yearly FLHs scheduling, respectively.}
		\label{tradeoff}
	\end{figure}

	The corresponding effect on the scheduling of hours by varying the angles of the inclined line in the proposed methodology is examined for the Danish dataset for the year 2018. The angle is varied at the rate of $x \in \{0^\circ, 15^\circ, 30^\circ, 45^\circ, 60^\circ, 75^\circ, 90^\circ\}$. Also, the hours are scheduled for daily, monthly, and yearly FLHs. Figure \ref{tradeoff} (a), (b), and (c) are the observed changes in the CO$_2$ intensity and electricity prices for 6000 FLHs with respect to change in the inclination angle of the line for daily, monthly and yearly FLHs scheduling, respectively. In these figures, 6000 FLHs are scheduled yearly, that is 500 hours scheduled monthly, and approximately 16 hours scheduled daily. The results for an angle of 45 degrees are highlighted with a big black dot. This is the case of equal weight between the electricity price and CO$_2$ intensity.
	
	It is clear from all sub figures that the price decreases and the CO$_2$ intensity increases with increasing angle and vice versa. This corresponds to the increasing (decreasing) angle weighting the electricity price (CO$_2$ intensity) higher. This leads to a trade-off between the two as it is not possible to obtain the lowest price and CO$_2$ intensity at the same time. Another general pattern is that as the scheduling horizon increases from daily to monthly to yearly, the variation in price and CO$_2$ intensity for different angles increases dramatically.
	
	For the daily scheduling, we see only small changes in price when changing the angle. However, the angle has a larger effect on the intensity. For the monthly scheduling, there seems to be a plateau in the intensity for angles between 30 degrees and 60 degrees. This opens an opportunity for weighting the price higher with an angle of 30 degrees and get about the same low intensity as if the angle was at 30 degrees. The lines for price and intensity diverge extremely when it comes to yearly scheduling. This means that any choice of angle in this case is a trade-off between the two.
	
	While the minima and maxima of the price and CO$_2$ lines get more extreme with increasing scheduling horizon, the values for the midpoint at 45 degrees decrease. From daily to monthly scheduling the price and CO$_2$ intensity decrease by 5.7 and 2.9 \%, respectively. From daily to yearly scheduling the price and CO$_2$ intensity decrease by 14.3 and 7.1 \%, respectively. This shows that with a longer scheduling horizon, it is on average possible to find more hours of simultaneous low price and CO$_2$ intensity than it is with only a daily horizon.

	\clearpage
	\section{Conclusions}
	With the increasing shares of variable renewable energy sources in the electricity system flexibility becomes increasingly important. Power-to-X assets are a great source of this flexibility. However, for these plants to be economic viable it is necessary to have proper scheduling mechanisms in place. In addition to considering market prices, these scheduling methods should also consider the CO$_2$ intensity of electricity. This ensures that the flexibility provided to the system by Power-to-X is supporting the integration of larger shares of variable renewable energy sources.
	
	This study has introduced a graphical approach to schedule Power-to-X plants while optimizing for minimum spot market electricity prices, minimum CO$_2$ intensity, or a mix between the two.
	
	In a simulation study using historical prices and CO$_2$ intensity for four different countries, it is found that there is a general tendency for the price and CO$_2$ intensity to increase with the number of full load hours scheduled. This is expected since more hours of operation means less flexibility in scheduling them. Another general pattern is that the price and CO$_2$ intensity tend to decrease with increasing scheduling horizon.
	
	The choice of optimization objective in scheduling has a huge impact on the resulting price and CO$_2$ emissions but the effect decreases fast with an increasing number of hours scheduled per year.
		
	Investigating the trade-off between optimizing for price or CO$_2$ intensity shows that it is indeed a trade-off: it is not possible to obtain the lowest price and CO$_2$ intensity at the same time. The potential for obtaining a low price or low CO$_2$ intensity increases with increasing scheduling horizon, but at the same time the trade-off becomes increasingly pronounced.
	
	There are two reasons for the simplicity of the graphical approach introduced in this study. Firstly, it is easy to explain to stakeholders without requiring deep mathematical knowledge. Secondly, it requires little effort to implement it in the operation of Power-to-X assets.
	
	In future work, we will explore more advanced methods that take the control constraints of the Power-to-X asset into account. These constraints include ramp rate limits and minimum running time after startup. Including such constraints are necessary for the scheduling method to be applicable to a wide range of Power-to-X assets.

	\section*{References}
	\bibliography{mybibfile}

\begin{thebibliography}{10}
\expandafter\ifx\csname url\endcsname\relax
  \def\url#1{\texttt{#1}}\fi
\expandafter\ifx\csname urlprefix\endcsname\relax\def\urlprefix{URL }\fi
\expandafter\ifx\csname href\endcsname\relax
  \def\href#1#2{#2} \def\path#1{#1}\fi

\bibitem{a1}
C.~J. Rhodes, {The 2015 Paris climate change conference: COP21}, Science
  Progress 99~(1) (2016) 97--104.

\bibitem{a2}
V.~Prakash, S.~Ghosh, K.~Kanjilal, Costs of avoided carbon emission from
  thermal and renewable sources of power in india and policy implications,
  Energy (2020) 117522.

\bibitem{a3}
N.~Bokde, A.~Feij{\'o}o, D.~Villanueva, K.~Kulat, A review on hybrid empirical
  mode decomposition models for wind speed and wind power prediction, Energies
  12~(2) (2019) 254.

\bibitem{a4}
S.~K. Mangla, S.~Luthra, S.~Jakhar, S.~Gandhi, K.~Muduli, A.~Kumar, A step to
  clean energy-sustainability in energy system management in an emerging
  economy context, Journal of Cleaner Production 242 (2020) 118462.

\bibitem{a5}
V.~Kolev, I.~Draganova-Zlateva, D.~Gospodinova, Trends in energy efficiency and
  co 2 emissions according to bulgarian national energy efficiency plan, in:
  2019 11th Electrical Engineering Faculty Conference (BulEF), IEEE, 2019, pp.
  1--6.

\bibitem{a6}
A.~D. Mills, T.~Levin, R.~Wiser, J.~Seel, A.~Botterud, Impacts of variable
  renewable energy on wholesale markets and generating assets in the united
  states: A review of expectations and evidence, Renewable and Sustainable
  Energy Reviews 120 (2020) 109670.

\bibitem{a7}
T.~Tsuji, H.~Bae, J.~Qi, Supply and demand balance control based on balancing
  power market, in: Economically Enabled Energy Management, Springer, 2020, pp.
  33--58.

\bibitem{a8}
X.~Yao, B.~Yi, Y.~Yu, Y.~Fan, L.~Zhu, Economic analysis of grid integration of
  variable solar and wind power with conventional power system, Applied Energy
  264 (2020) 114706.

\bibitem{a9}
J.~C. Koj, C.~Wulf, P.~Zapp, {Environmental impacts of power-to-X systems-A
  review of technological and methodological choices in Life Cycle
  Assessments}, Renewable and Sustainable Energy Reviews 112 (2019) 865--879.

\bibitem{a10}
J.~Burre, D.~Bongartz, L.~Br{\'e}e, K.~Roh, A.~Mitsos, {Power-to-X: Between
  Electricity Storage, e-Production, and Demand Side Management}, Chemie
  Ingenieur Technik 92~(1-2) (2020) 74--84.

\bibitem{a11}
E.~Panos, T.~Kober, A.~Wokaun, Long term evaluation of electric storage
  technologies vs alternative flexibility options for the swiss energy system,
  Applied Energy 252 (2019) 113470.

\bibitem{a12}
M.~Bailera, P.~Lisbona, L.~M. Romeo, S.~Espatolero, {Power to Gas projects
  review: Lab, pilot and demo plants for storing renewable energy and CO2},
  Renewable and Sustainable Energy Reviews 69 (2017) 292--312.

\bibitem{a13}
H.~Blanco, A.~Faaij, A review at the role of storage in energy systems with a
  focus on power to gas and long-term storage, Renewable and Sustainable Energy
  Reviews 81 (2018) 1049--1086.

\bibitem{a14}
J.~C. Koj, C.~Wulf, P.~Zapp, {Environmental impacts of power-to-X systems-A
  review of technological and methodological choices in Life Cycle
  Assessments}, Renewable and Sustainable Energy Reviews 112 (2019) 865--879.

\bibitem{a15}
Z.~Chehade, C.~Mansilla, P.~Lucchese, S.~Hilliard, J.~Proost, {Review and
  analysis of demonstration projects on power-to-X pathways in the world},
  International Journal of Hydrogen Energy 44~(51) (2019) 27637--27655.

\bibitem{a16}
B.~Rego~de Vasconcelos, J.-M. Lavoie, {Recent advances in Power-to-X technology
  for the production of fuels and chemicals}, Frontiers in Chemistry 7 (2019)
  392.

\bibitem{a17}
B.~Aust, A.~Horsch, Negative market prices on power exchanges: Evidence and
  policy implications from germany, The Electricity Journal 33~(3) (2020)
  106716.

\bibitem{a18}
L.~Janke, S.~McDonagh, S.~Weinrich, J.~Murphy, D.~Nilsson, P.-A. Hansson,
  {\AA}.~Nordberg, {Optimizing power-to-H2 participation in the Nord Pool
  electricity market: Effects of different bidding strategies on plant
  operation}, Renewable Energy.

\bibitem{a19}
Z.~Chehade, C.~Mansilla, P.~Lucchese, S.~Hilliard, J.~Proost, {Review and
  analysis of demonstration projects on power-to-X pathways in the world},
  International Journal of Hydrogen Energy 44~(51) (2019) 27637--27655.

\bibitem{a20}
C.~Schnuelle, J.~Thoeming, T.~Wassermann, P.~Thier, A.~von Gleich,
  S.~Goessling-Reisemann, {Socio-technical-economic assessment of power-to-X:
  Potentials and limitations for an integration into the German energy system},
  Energy Research \& Social Science 51 (2019) 187--197.

\bibitem{a21}
T.~Kober, C.~Bauer, C.~Bach, M.~Beuse, G.~Georges, M.~Held, S.~Heselhaus,
  P.~Korba, L.~K{\"u}ng, A.~Malhotra, et~al., {Perspectives of Power-to-X
  technologies in Switzerland}, Tech. rep., ETH Zurich (2019).

\bibitem{a22}
S.~Nielsen, I.~R. Skov, Investment screening model for spatial deployment of
  power-to-gas plants on a national scale--a danish case, International Journal
  of Hydrogen Energy 44~(19) (2019) 9544--9557.

\bibitem{a23}
M.~T{\"a}htinen, T.~Sihvonen, J.~Savolainen, R.~Weiss, {Interim H2 storage in
  power to gas process: Dynamic unit process modelling and dynamic
  simulations}, in: 10th International Renewable Energy Storage, IRES 2016,
  EUROSOLAR The European Association for Renewable Energy, 2016.

\bibitem{a25}
Y.~Li, W.~Liu, M.~Shahidehpour, F.~Wen, K.~Wang, Y.~Huang, Optimal operation
  strategy for integrated natural gas generating unit and power-to-gas
  conversion facilities, IEEE Transactions on Sustainable Energy 9~(4) (2018)
  1870--1879.

\bibitem{a26}
J.~Yang, N.~Zhang, Y.~Cheng, C.~Kang, Q.~Xia, {Modeling the operation mechanism
  of combined P2G and gas-fired plant with CO2 recycling}, IEEE Transactions on
  Smart Grid 10~(1) (2018) 1111--1121.

\bibitem{a24}
H.~Khani, H.~E.~Z. Farag, Optimal day-ahead scheduling of power-to-gas energy
  storage and gas load management in wholesale electricity and gas markets,
  IEEE Transactions on Sustainable Energy 9~(2) (2017) 940--951.

\bibitem{b1}
A.~Kikuchi, M.~Ito, Y.~Hayashi, Scheduling method of wind power generation for
  electricity market using state-of-charge transition and forecast error,
  Journal of International Council on Electrical Engineering 9~(1) (2019)
  123--132.

\bibitem{b2}
T.~Windler, J.~Busse, J.~Rieck, One month-ahead electricity price forecasting
  in the context of production planning, Journal of Cleaner Production 238
  (2019) 117910.

\bibitem{b3}
Y.~Cao, J.~Wei, C.~Li, B.~Zhou, L.~Huang, G.~Feng, H.~Yang, Optimal operating
  control strategy for biogas generation under electricity spot market, The
  Journal of Engineering 2019~(18) (2019) 5183--5186.

\bibitem{b4}
N.~Bokde, B.~Tranberg, G.~B. Andresen, Short-term {CO2} emissions forecasting
  based on decomposition approaches, arXiv preprint arXiv:2003.10868.

\bibitem{rad2020optimal}
A.~M. Rad, T.~Barforoushi, Optimal scheduling of resources and appliances in
  smart homes under uncertainties considering participation in spot and
  contractual markets, Energy 192 (2020) 116548.

\bibitem{zhao2019multi}
J.~Zhao, T.~Zheng, E.~Litvinov, A multi-period market design for markets with
  intertemporal constraints, IEEE Transactions on Power Systems.

\bibitem{transparency}
{ENTSO-E Transparency Platform}, \url{https://transparency.entsoe.eu}.

\bibitem{tranberg2019real}
B.~Tranberg, O.~Corradi, B.~Lajoie, T.~Gibon, I.~Staffell, G.~B. Andresen,
  Real-time carbon accounting method for the {European} electricity markets,
  Energy Strategy Reviews 26 (2019) 100367.
\newblock \href {http://dx.doi.org/10.1016/j.esr.2019.100367}
  {\path{doi:10.1016/j.esr.2019.100367}}.

\bibitem{Tranberg2015}
B.~Tranberg, A.~Thomsen, R.~Rodriguez, G.~Andresen, M.~Sch\"afer, M.~Greiner,
  \href{http://stacks.iop.org/1367-2630/17/i=10/a=105002}{{Power flow tracing
  in a simplified highly renewable European electricity networks}}, New Journal
  of Physics 17 (2015) 105002.
\newblock \href {http://dx.doi.org/10.1088/1367-2630/17/10/105002}
  {\path{doi:10.1088/1367-2630/17/10/105002}}.
\newline\urlprefix\url{http://stacks.iop.org/1367-2630/17/i=10/a=105002}

\bibitem{ecoinvent}
G.~Wernet, C.~Bauer, B.~Steubing, J.~Reinhard, E.~Moreno-Ruiz, B.~Weidema,
  \href{https://doi.org/10.1007/s11367-016-1087-8}{The ecoinvent database
  version 3 (part {I}): overview and methodology}, The International Journal of
  Life Cycle Assessment 21~(9) (2016) 1218--1230.
\newblock \href {http://dx.doi.org/10.1007/s11367-016-1087-8}
  {\path{doi:10.1007/s11367-016-1087-8}}.
\newline\urlprefix\url{https://doi.org/10.1007/s11367-016-1087-8}

\bibitem{bokde2020forecasttb}
N.~D. Bokde, Z.~M. Yaseen, G.~B. Andersen, {ForecastTB—An R Package as a
  Test-Bench for Time Series Forecasting—Application of Wind Speed and Solar
  Radiation Modeling}, Energies 13~(10) (2020) 2578.

\end{thebibliography}
	
	\appendix
	\section{Tables}
		\begin{table}[p]
		\centering
		\caption{Percentage improvements in electricity prices and CO$_2$ intensity values at different FLHs and time scale scheduling in Denmark with the proposed trade-off methodology. (values are rounded at the decimal point).}
		\label{T1}
		\begingroup\fontsize{9pt}{10pt}\selectfont
		\makebox[\columnwidth]{\begin{tabular}{ccccccc}
				\hline
				FLH                   & Time scale               & Parameters & Best-case & Compromised-case & Trade-off   case & \%   Improvement \\ \hline
				\multirow{6}{*}{4000} & \multirow{2}{*}{Yearly}  & Intensity  & 236    & 328          & 245          & 25           \\
				&                          & Prices     & 33    & 44           & 34            & 21            \\ \cline{2-7}
				& \multirow{2}{*}{Monthly} & Intensity  & 253    & 313           & 279           & 10            \\
				&                          & Prices     & 36     & 41            & 38            & 8             \\ \cline{2-7}
				& \multirow{2}{*}{Daily}   & Intensity  & 285    & 333           & 322           & 3             \\
				&                          & Prices     & 38     & 44            & 40            & 9             \\ \hline
				\multirow{6}{*}{5000} & \multirow{2}{*}{Yearly}  & Intensity  & 252    & 326           & 258           & 20            \\
				&                          & Prices     & 36     & 44            & 37            & 17             \\ \cline{2-7}
				& \multirow{2}{*}{Monthly} & Intensity  & 268    & 317           & 289           & 8            \\
				&                          & Prices     & 38     & 42            & 39            & 8             \\ \cline{2-7}
				& \multirow{2}{*}{Daily}   & Intensity  & 294    & 331           & 315           & 4             \\
				&                          & Prices     & 40     & 44            & 41            & 8             \\ \hline
				\multirow{6}{*}{6000} & \multirow{2}{*}{Yearly}  & Intensity  & 268    & 328           & 273           & 16            \\
				&                          & Prices     & 38     & 45            & 39            & 14             \\ \cline{2-7}
				& \multirow{2}{*}{Monthly} & Intensity  & 282    & 319           & 301           & 5             \\
				&                          & Prices     & 39     & 43            & 41            & 4             \\ \cline{2-7}
				& \multirow{2}{*}{Daily}   & Intensity  & 301    & 330           & 319           & 3             \\
				&                          & Prices     & 41     & 45            & 42            & 6             \\ \hline
				\multirow{6}{*}{7000} & \multirow{2}{*}{Yearly}  & Intensity  & 286    & 329           & 289           & 12            \\
				&                          & Prices     & 40     & 45            & 41            & 10             \\ \cline{2-7}
				& \multirow{2}{*}{Monthly} & Intensity  & 297    & 321           & 311           & 3             \\
				&                          & Prices     & 41     & 44            & 43            & 3             \\ \cline{2-7}
				& \multirow{2}{*}{Daily}   & Intensity  & 311    & 330           & 317           & 4             \\
				&                          & Prices     & 43     & 45            & 44            & 1             \\ \hline
		\end{tabular}}
		\endgroup
	\end{table}

\begin{center}
	\begingroup\fontsize{9pt}{10pt}\selectfont
	\begin{longtable}{cccccc}
		\caption{Percentage improvements in electricity prices and CO$_2$ intensity values at 6000 FLHs for yearly scheduling in 26 areas with the proposed trade-off methodology. (values are rounded at the decimal point).} \label{T2}\\
		\hline
		Countries            & Parameter & Best case & Compromised case & Trade-off case & \% Improvement \\ \hline
		\multirow{2}{*}{DK-2} & Intensity & 268    & 328           & 273         & 16          \\
		& Prices    & 38     & 45            & 42          & 6           \\
		\multirow{2}{*}{DK-1} & Intensity & 189    & 276           & 197         & 28          \\
		& Prices    & 36     & 42            & 39          & 7           \\
		\multirow{2}{*}{AT}   & Intensity & 228 & 260         & 235      & 9           \\
		& Prices    & 37   & 42         & 40       & 5           \\
		\multirow{2}{*}{BE}   & Intensity & 266 & 356        & 269         & 24          \\
		& Prices    & 44  & 55         & 53        & 3           \\
		\multirow{2}{*}{BG}   & Intensity & 494 & 528         & 499      & 5           \\
		& Prices    & 59  & 69         & 63       & 9           \\
		\multirow{2}{*}{CH}   & Intensity & 431    & 490           & 442         & 9           \\
		& Prices    & 43     & 52            & 48          & 7           \\
		\multirow{2}{*}{CZ}   & Intensity & 602 & 690        & 618      & 10          \\
		& Prices    & 37  & 46         & 43       & 6           \\
		\multirow{2}{*}{DE}   & Intensity & 459 & 486         & 508       & -4          \\
		& Prices    & 35  & 39         & 41       & -5           \\
		\multirow{2}{*}{EE}   & Intensity & 872 & 948        & 918       & 3           \\
		& Prices    & 40  & 46         & 48       & -3           \\
		\multirow{2}{*}{ES}   & Intensity & 311 & 359        & 325      & 9          \\
		& Prices    & 51 & 55         & 54       & 2           \\
		\multirow{2}{*}{FI}   & Intensity & 298  & 322        & 301      & 6           \\
		& Prices    & 40  & 45         & 43       & 3           \\
		\multirow{2}{*}{FR}   & Intensity & 60  & 66          & 61       & 7           \\
		& Prices    & 41   & 43         & 42       & 1           \\
		\multirow{2}{*}{GB}   & Intensity & 291 & 315        & 296      & 6           \\
		& Prices    & 50   & 54         & 52       & 4           \\
		\multirow{2}{*}{GR}   & Intensity & 650 & 697        & 679      & 2           \\
		& Prices    & 55 & 59         & 58       & 1           \\
		\multirow{2}{*}{HU}   & Intensity & 409 & 422        & 445      & -5           \\
		& Prices    & 41  & 45         & 59       & -30           \\
		\multirow{2}{*}{IT}   & Intensity & 468 & 493        & 473      & 4           \\
		& Prices    & 52  & 59          & 56       & 6           \\
		\multirow{2}{*}{LT}   & Intensity & 253 & 289        & 258      & 10          \\
		& Prices    & 41   & 46          & 44       & 3           \\
		\multirow{2}{*}{LV}   & Intensity & 442 & 521        & 447      & 14          \\
		& Prices    & 41  & 50         & 47       & 5           \\
		\multirow{2}{*}{NL}   & Intensity & 352 & 442         & 361      & 18          \\
		& Prices    & 44  & 52          & 47       & 8           \\
		\multirow{2}{*}{NO}   & Intensity & 28  & 65         & 32       & 50          \\
		& Prices    & 39  & 45         & 43       & 4           \\
		\multirow{2}{*}{PL}   & Intensity & 899  & 925         & 937       & -1           \\
		& Prices    & 186  & 209          & 211      & -1           \\
		\multirow{2}{*}{PT}   & Intensity & 329 & 358          & 366      & -2          \\
		& Prices    & 52  & 54         & 57       & -5           \\
		\multirow{2}{*}{RO}   & Intensity & 354   & 380          & 394       & -3           \\
		& Prices    & 163  & 201          & 201      & 0           \\
		\multirow{2}{*}{RS}   & Intensity & 816 & 863        & 833      & 3           \\
		& Prices    & 40  & 47         & 44       & 6           \\
		\multirow{2}{*}{SI}   & Intensity & 317 & 359        & 331      & 7          \\
		& Prices    & 41  & 49         & 43       & 11          \\
		\multirow{2}{*}{SK}   & Intensity & 384 & 409        & 419      & -2          \\
		& Prices    & 38  & 44         & 46       & -3         \\ \cline{2-6}
	\end{longtable}
	\endgroup
\end{center}

	\section{Performance of the proposed trade-off methodology for other countries}
	\label{app}
\begin{figure}[p]
		\centering{\includegraphics[width=\textwidth]{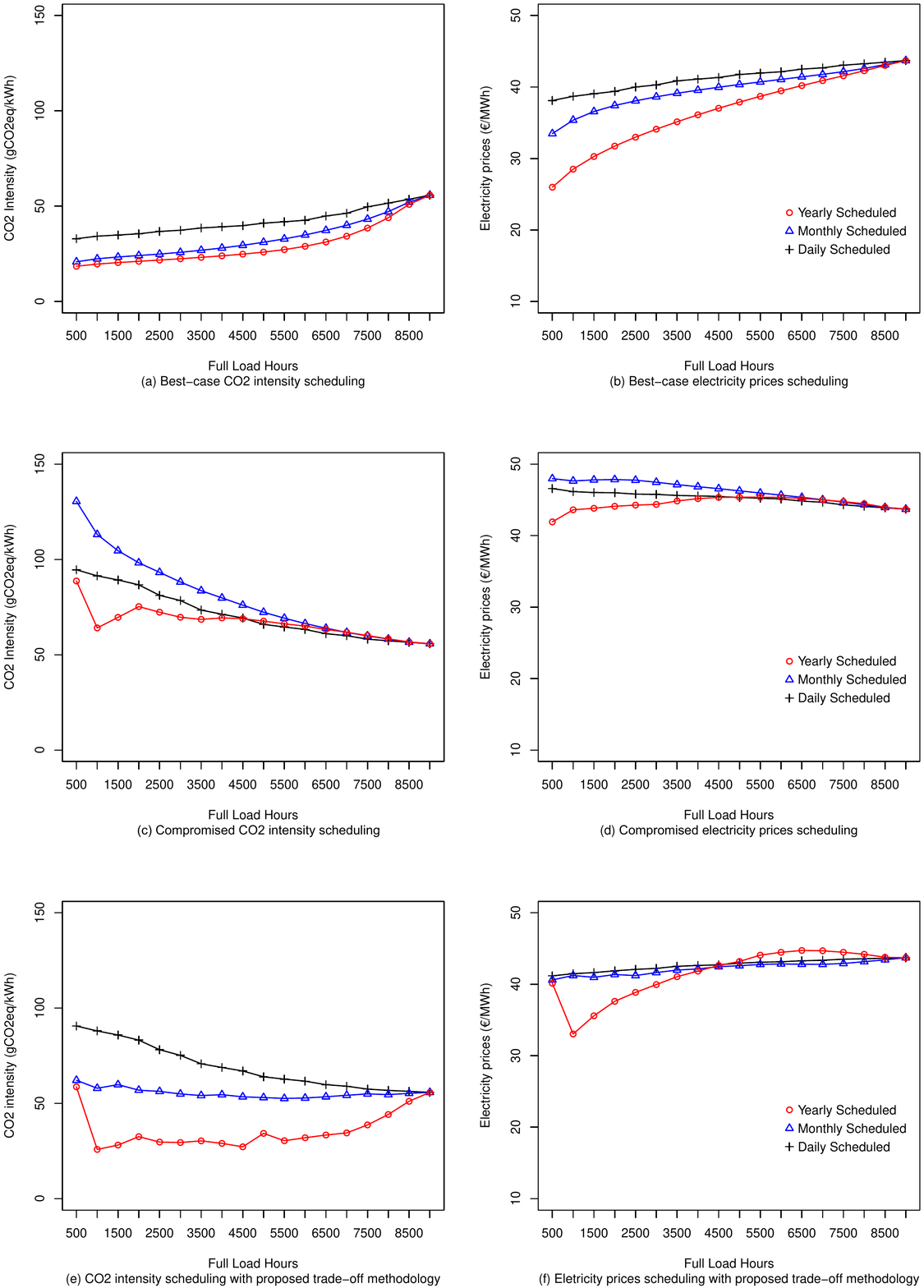}}
		\caption{Full load hours scheduling in Norway.}
		\label{Fig6}
	\end{figure}

\begin{figure}[p]
		\centering{\includegraphics[width=\textwidth]{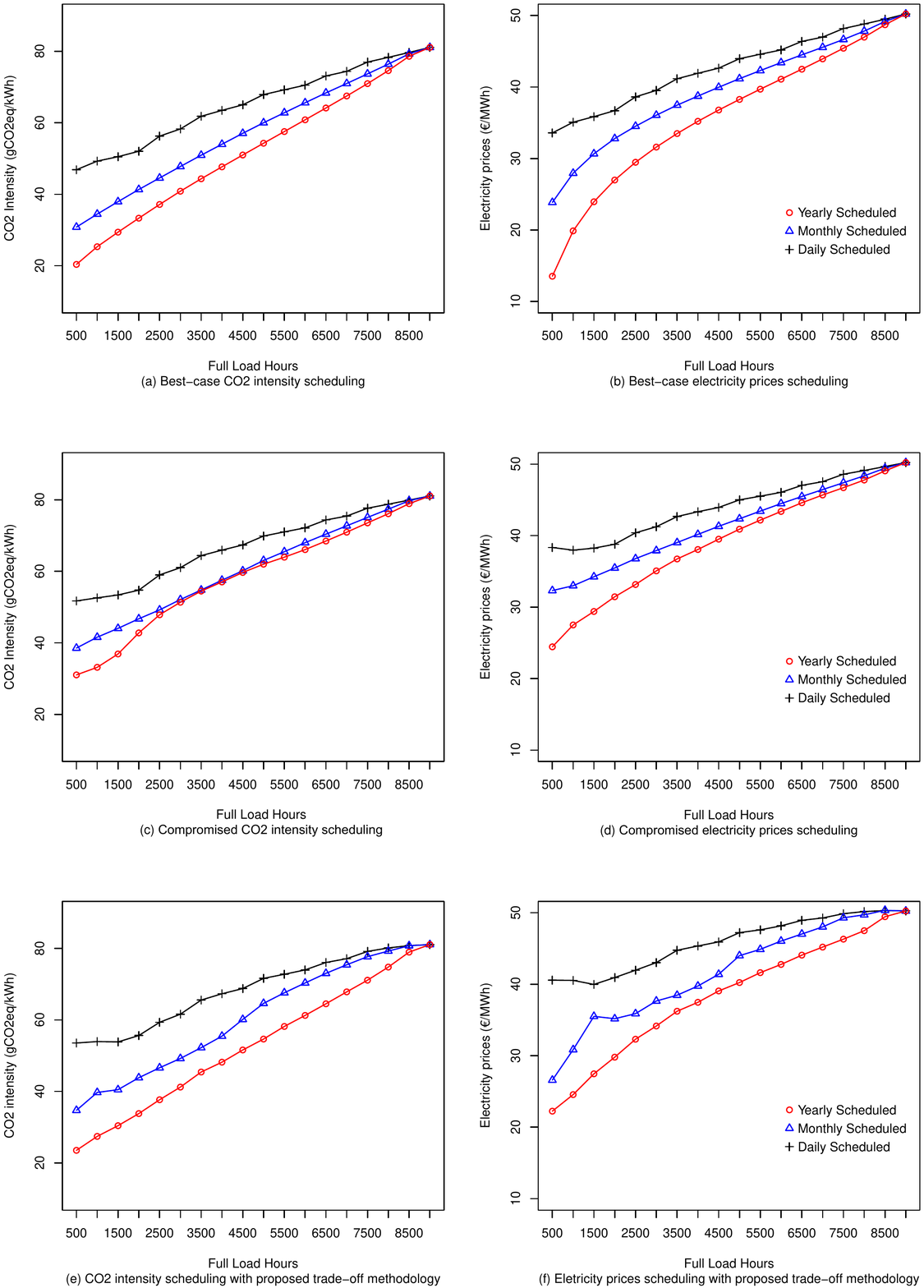}}
		\caption{Full load hours scheduling in France.}
		\label{Fig7}
	\end{figure}

\begin{figure}[p]
		\centering{\includegraphics[width=\textwidth]{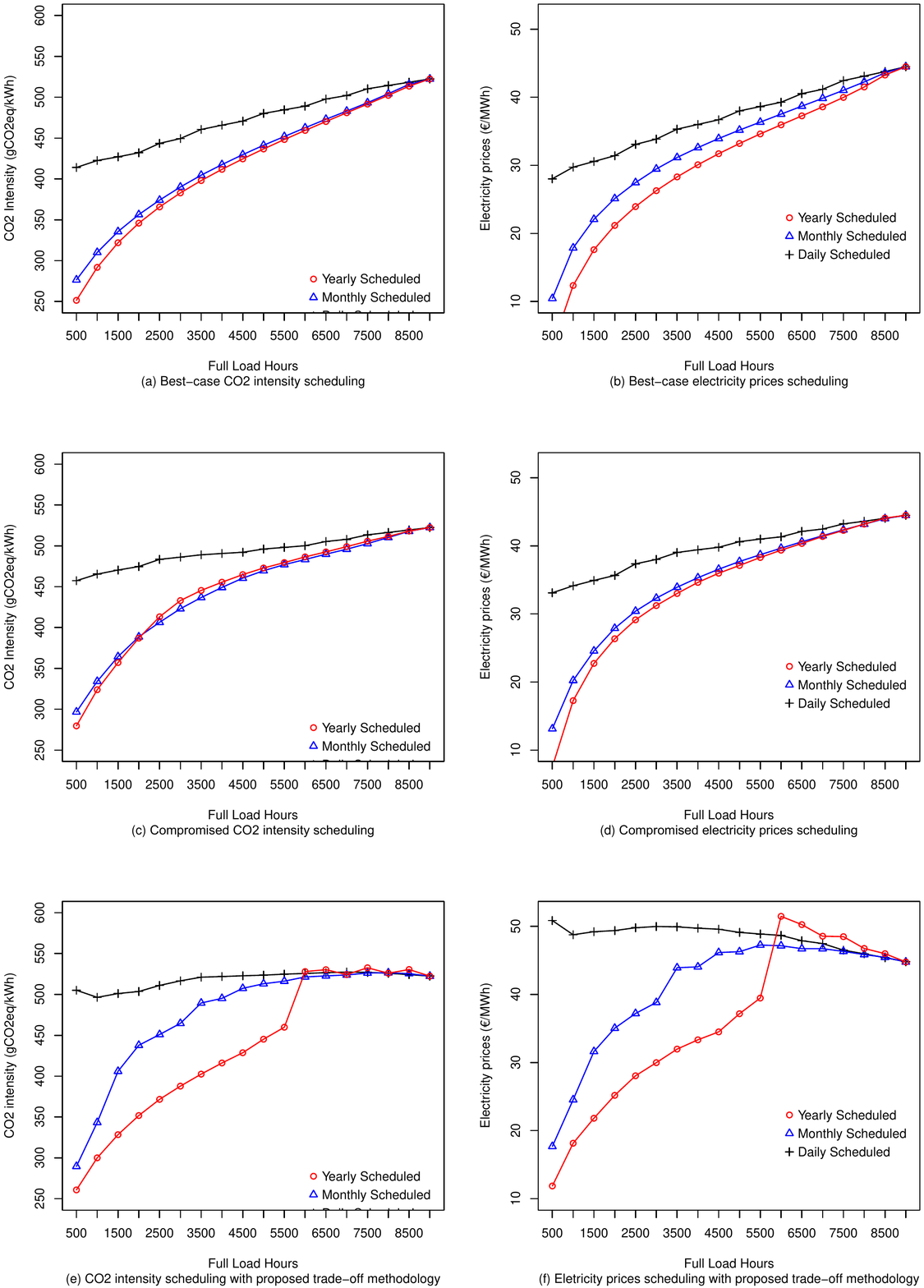}}
		\caption{Full load hours scheduling in Germany.}
		\label{Fig8}
	\end{figure}

	This section have the Figures \ref{Fig6}, \ref{Fig7} and \ref{Fig8}, which show the performance of the proposed trade-off methodology in Norway, France, and Germany, respectively.
	\newgeometry{textwidth=6in}
	\restoregeometry
	
\end{document}